\documentclass[letterpaper]{emulateapj}
\usepackage{apjfonts}
\usepackage{amsmath}
\usepackage{natbib}
\usepackage[colorlinks=true,linkcolor=blue,citecolor=blue]{hyperref}
\usepackage[normalem]{ulem}

\begin{document}

\newcommand{\beq}{\begin{equation}}
\newcommand{\eeq}{\end{equation}}
\newcommand{\beqar}{\begin{eqnarray}}
\newcommand{\eeqar}{\end{eqnarray}}
\newcommand{\e}{\varepsilon}
\newcommand{\rt}{r_{\rm t}}
\newcommand{\rs}{r_{\rm s}}
\newcommand{\Mbh}{M_{\rm bh}}
\newcommand{\rp}{r_{\rm p}}
\newcommand{\rb}{r_{\rm b}}
\newcommand{\tdyn}{t_{\rm dyn}}
\newcommand{\tfb}{t_{\rm fb}}
\newcommand{\Jc}{J_{\rm c}}
\newcommand{\Jlc}{J_{\rm lc}}
\newcommand{\tr}{t_{\rm r}}
\newcommand{\tJ}{t_{\rm J}}
\newcommand{\Flc}{F_{\rm lc}}

\shortauthors{MacLeod, Guillochon, \& Ramirez-Ruiz}

\title{The Tidal Disruption of Giant Stars and Their  Contribution to the Flaring Supermassive Black Hole Population} 
\shorttitle{Tidal Disruption of Giant Stars}
\author{Morgan MacLeod, James Guillochon, and Enrico Ramirez-Ruiz}
\affil{Department of Astronomy and
  Astrophysics, University of California, Santa Cruz, CA
  95064}
\email{mmacleod, jfg, enrico @ucolick.org}
   
 \begin{abstract}
Sun-like stars are thought to be regularly disrupted by supermassive black holes (SMBHs) within galactic nuclei. Yet, as stars evolve off the main sequence their vulnerability to tidal disruption  increases drastically as they develop a bifurcated structure consisting of  a dense core and a tenuous envelope. Here we present the first hydrodynamic simulations of the tidal disruption of giant stars and show that the core has a substantial influence on the star's ability to survive the encounter. Stars with more massive cores retain large fractions of their envelope mass, even in deep encounters. Accretion flares resulting from the disruption of giant stars should last for tens to hundreds of years. Their characteristic signature in transient searches would not be the $t^{-5/3}$ decay typically associated with tidal disruption events, but a correlated rise over many orders of magnitude in brightness on months to years timescales. We calculate the relative disruption rates of stars of varying evolutionary stages in typical galactic centers, then use our results to produce Monte Carlo realizations of the expected flaring event populations. 
We find that the demographics of tidal disruption flares are strongly dependent on both stellar and black hole mass, especially near the limiting SMBH mass scale of $\sim 10^8 M_\odot$. 
At this black hole mass, we predict a sharp transition in the SMBH flaring diet beyond which all observable disruptions arise from evolved stars, accompanied by a dramatic cutoff in the overall tidal disruption flaring rate. Black holes less massive than this limiting mass scale will show observable flares from both main sequence and evolved stars, with giants contributing up to 10\% of the event rate. The relative fractions of stars disrupted at different evolutionary states can constrain the properties and distributions of stars in galactic nuclei other than our own.
\end{abstract}

\maketitle

%
%
\section{Introduction}

Quasars are rapidly growing black holes lit up by the gas they accrete. They are the most dramatic manifestation of
the more general phenomenon of active galactic nuclei (AGN) and they are among the most energetic objects in the universe.
Several billion years after the Big Bang, the universe went through a quasar era when highly-luminous AGN were a standard feature of most massive galaxies  \citep{Kormendy:1995hc,Richstone:1998wk}. 
Since that time, AGN have been dying out and the only  activity that still occurs in many nearby galactic nuclei is weak  \citep{Schawinski:2010bl}.
All that should remain in the centers of local galaxies are the remnants of quasar-era exponential growth: quiescent SMBHs, now dim and starved of fuel \citep{Ho:2008kz}. 
It is, therefore, not surprising that definitive conclusions about the presence of local SMBHs are typically drawn from very nearby galaxies with little to no AGN activity \citep{Gebhardt2000,Ferrarese2000}. The centers of these galaxies are well resolved, revealing the region where the black hole dominates the stellar dynamics. 

The question then arises of whether the postulated presence of SMBHs lurking in the centers of most galaxies is consistent with the apparent quiescence of their nuclei. 
Quiescent black holes are, in fact, nearly black; they may only be lit up by the luminance of accreting matter. We do not directly know how much gas there is near most SMBHs, and there is no a priori reason why nuclear regions should be swept completely clean of gas. 
The distribution of stars in dense clusters that surround SMBHs, on the other hand, is much better constrained.  These densely packed stars trace complicated and wandering orbits under the combined influence of all the other stars and the black hole itself. If a star wanders too close to the black hole it is violently ripped apart by the hole's tidal field \citep[eg,][]{Hills:1975kh,Frank:1978wx,Rees1988}. About half of the debris of tidal disruption eventually falls back  and accretes onto the SMBH. This accretion powers a flare which is a definitive sign of the presence of an otherwise quiescent SMBH and a powerful diagnostic of its properties \citep{Rees1988}. 
Tidal disruption events are expected to be relatively rare, on the order of one per $\sim10^4$ years per galaxy \citep[e.g.][]{Magorrian:1999fd,Wang:2004jy}.  Depending primarily on the structural properties of the disrupted star, an ultra-luminous transient  signal could persist steadily for months to at most  tens of years;  thereafter the flare would rapidly fade. 
 In a given galaxy, this luminous flaring activity would then have a short duty cycle. Quiescent SMBHs should greatly outnumber active ones. Observational constraints would not, therefore, be stringent until we had observed enough candidates to constitute a proper ensemble average. However, the long decay tails of these flares may account for an appreciable fraction of the total low-luminosity AGN activity in the local universe \citep{Milosavljevic:2006jj}.

The critical pericenter distance for tidal disruption is the tidal radius,
\beq\label{rt}
\rt \equiv \left( \frac{M_{\rm bh}}{M_\ast}\right)^{1/3} R_\ast,
\eeq
where $M_\ast$ and $R_\ast$ are the stellar mass and radius, and $M_{\rm bh}$ is the black hole mass \citep{Hills:1975kh}. The tidal radius is larger than the black hole's horizon, $\rs = 2 G M_{\rm bh} /c^2$, for solar type stars as long as the black hole is less massive than about $10^8$ solar masses. In encounters with more massive black holes, solar type stars may pass the horizon undisrupted and are effectively swallowed whole. Such events would leave little electromagnetic signature \citep[although a portion of the gravitational wave signature would remain,][]{Kobayashi:2004kq}. However, for a given stellar and black hole mass, evolved stars have larger tidal radii than main sequence (MS) stars and are therefore more vulnerable to tidal disruption. Furthermore, giant branch stars are the only stars that can produce observable tidal disruption flares in encounters with the most massive black holes $\gtrsim 10^8 M_\odot$. 

Motivated by these facts, we examine the importance of stellar evolution in the context of tidal disruption. A sun-like star spends the majority of its   lifetime on the MS, $\tau_{\rm ms}\sim 10^{10} $ years, followed by a relatively brief period of post-MS (giant branch) evolution, $\tau_{\rm g} \sim 10^8$ years, once its central supply of hydrogen fuel is exhausted. As nuclear reactions slow,   the stellar core loses pressure support, and, in approximately a thermal diffusion time, the star ascends the giant branch as its outer layers expand in response to the core's collapse. 
Giant-branch stars are much less common than MS stars in a typical stellar population due to the ratio of their  lifetimes, which for a solar mass star is $\tau_{\rm g}/\tau_{\rm ms} \sim 10^{-2}$. 
 However, their large radii imply that they are exceptionally vulnerable to tidal disruption during these brief periods. The contribution of giant stars to the tidal disruption event rate and the luminosity function of AGN will depend on the competing effects of their enhanced cross-section and their relative rarity.

 Understanding the details of giant star disruption events and their contribution to the SMBH flaring population is the focus of this work. We use several methods to study this problem. 
In Section \ref{sec:SEandTD}, we discuss the calculation of detailed stellar evolution models and outline the importance of stellar evolution in the context of tidal disruption.
The non-linear dynamics of the encounters themselves must be understood through hydrodynamic simulations; this is particularly true for post-MS stars which are not well described by a simple single-polytrope model.
In Section \ref{sec:hydro}, we describe how we derive giant star initial models from our stellar evolution calculations, our methods of hydrodynamic simulation, and the results of our simulations of close encounters between giant-branch stars and SMBHs. 
In Section \ref{sec:lc}, we calculate the rates of tidal disruption that result from the two-body relaxation driven random walk of nuclear cluster stars in angular momentum space. We focus on the relative rates of disruption of stars in different evolutionary states. 
In Section \ref{sec:disc}, we combine our stellar evolution, hydrodynamic, and rate calculations of Sections \ref{sec:SEandTD} - \ref{sec:lc} and present Monte Carlo realizations of flaring events. We discuss the demographics of tidal disruption-powered flaring events as a function of black hole mass, the contribution of giant stars to the luminosity function of local AGN, and the detection of flares due to the disruption of giant stars.

%
%

\section{Stellar Evolution in the Context of Tidal Disruption}\label{sec:SEandTD}

Stellar evolution naturally enriches the physics of  tidal disruption and leads to a large diversity of accretion  flare events. The effects of stellar evolution are most pronounced as stars evolve off the MS. Their radius can change by orders of magnitude, they often suffer some degree of mass loss, and the density contrast  between the core and the envelope dramatically increases. In this section, we will outline the basic characteristics of post-MS stellar evolution and provide some intuition for how the capture rates and properties of the tidal disruption events might change as a stars evolve off the MS.  

\subsection{Stellar evolution models}\label{sec:SEandTD:methods}

To capture the intricacies of post-MS stellar evolution we use the open source Modules for Experiments in Stellar Astrophysics (MESA) stellar evolution code \citep[version 3290,][]{Paxton:2011jf}. MESA solves the stellar evolution equations within the Lagrangian formalism, closed by a tabulated equation of state which blends OPAL, SCVH, PC, and Helmholz equations of state to treat a wide range of fluid conditions and ionization states \citep{Paxton:2011jf,Rogers:2002cr,Saumon:1995bu,Potekhin:2010fu,Timmes:2000wm}. MESA handles short evolutionary periods  (for example, a helium core flash) by switching to a hydrodynamic solution which allows non-zero velocities at cell interfaces. This capability gives MESA the ability to efficiently evolve stars from their pre-MS collapse all the way to the formation of a white dwarf. For our purposes this is essential because it allows us to capture the entire lifetime of stars. 

We compute models of solar metallically stars in the mass range of $0.95 -5 M_\odot$. To do so, we employ an extension to MESA's basic hydrogen and helium nuclear burning network called \texttt{agb.net}.
Stellar mass loss is modeled for red giant (RG) stars using Reimers's formula,
\beq
\dot M_{\rm R} = 4 \times 10^{-13} \eta_{\rm R} \left(\frac{L_\ast}{L_\odot}\right) \left(\frac{R_\ast}{R_\odot}\right) \left(\frac{M_\ast}{M_\odot}\right)^{-1} \ M_\odot \ {\rm yr}^{-1},
\eeq
with $\eta_{\rm R}  = 0.5$ \citep{Reimers:1975vw}. Mass loss on the asymptotic giant branch (AGB) is given by the Bl\"ocker formula,
\beq
\dot M_{\rm B} = 1.932 \times 10^{-21} \eta_{\rm B} \left(\frac{L_\ast}{L_\odot}\right)^{3.7} \left(\frac{R_\ast}{R_\odot}\right) \left(\frac{M_\ast}{M_\odot}\right)^{-3.1} \ M_\odot \ {\rm yr}^{-1},
\eeq
with $\eta_{\rm B}= 0.1$ \citep{Bloecker:1995ui}. A switch between the wind schemes is triggered when the central hydrogen and helium are depleted. The efficiencies of wind mass loss rates are relatively uncertain \citep[e.g.][]{Habing:2003uy}, and they affect some  properties of giant stars. The efficiencies $\eta_{\rm R} = 0.5$ and $\eta_{\rm B} = 0.1$  are typical values used in the fiducial $1M_\odot$ models presented with the MESA code \citep{Paxton:2011jf}. 

\subsection{Post-MS stellar evolution} 

\begin{figure*}[tbp]
\begin{center}
\includegraphics[width=0.85\textwidth]{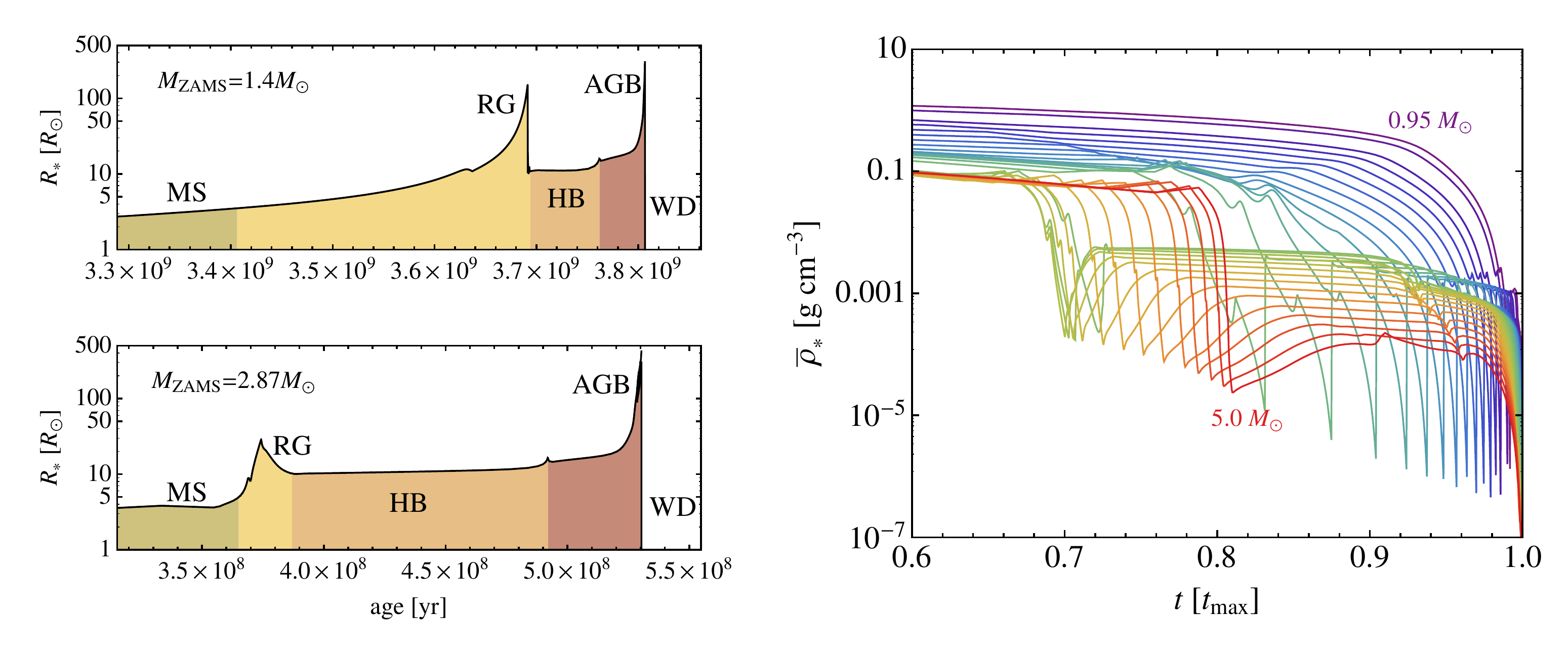}
\caption{Radius (left) and mean density (right) evolution of post-MS stars. In the left panel, stages are marked as: main sequence (MS), red giant (RG), horizontal branch (HB), and asymptotic giant branch (AGB).  In the right panel, we plot the mean density of stars between 0.95 and 5 $M_\odot$ as a function of fractional lifetime. The mean density describes the stars' vulnerability to tidal disruption \eqref{rt}. Periods of low density occupy an increasing portion of the lifetime of more massive stars.  }
\label{fig:EvolveStages}
\end{center}
\end{figure*}

Stars exhaust their central reserves of hydrogen fuel in about $\tau_{\rm ms} \sim 10^{10} (M_\ast/M_\odot)^{-2.9}$ years \citep{Hansen:2004ul}. After this, they begin to evolve off the MS. Nuclear reactions slow in the stellar core, which contracts under the influence of its self-gravity as it loses pressure support in approximately a thermal diffusion timescale. Hydrogen burns to helium off-center in a thick shell that continues to add mass to the now inert helium core \citep{Iliadis:2007ta}. The star's structure reacts to the core's contraction through an expansion of the star's outer layers. During this phase, the star's luminosity increases greatly, while its effective temperature decreases causing it to ascend the RG branch in the Hertzsprung-Russell (HR) diagram. The star develops an increasingly bifurcated structure with a dense, radiative core and a convective envelope. In stars less than about $2.2 M_\odot$, the helium core collapses to the point where it becomes  degenerate. When helium burning eventually ignites it does so in degenerate material, causing a dynamic event known as the helium core flash. Above about $2.2 M_\odot$, helium ignition occurs off-center in non-degenerate material, and the star does not reach as extreme radii while on the RG branch \citep{Maeder:2009cd}.

Following helium ignition, the core expands and the star stably burns helium to carbon and oxygen for a period of about $\tau_{\rm hb} \sim 10^8$ years  along the horizontal branch (HB). During this time the core mass grows significantly as hydrogen continues to burn to helium in a shell surrounding the core \citep{Maeder:2009cd}. Finally, the star may ascend to the tip of the giant branch one final time as its carbon oxygen core becomes degenerate. Stars above (about) $9M_\odot$ will go through  additional burning phases  \citep{Maeder:2009cd}. For the intermediate mass stars considered here, at the tip of the AGB, the stars' luminosity becomes so great that the bulk of the envelope is driven off via a combination of strong winds and thermal pulses, eventually exposing the bare carbon-oxygen core -- a proto-white dwarf \citep{Habing:2003uy}. 

In the left panel of Figure \ref{fig:EvolveStages}, we show typical evolutionary stages for stars with  $1.4 M_\odot$ and $2.87 M_\odot$ zero age MS (ZAMS) mass. For simplicity we define here the transition from the MS to the sub-giant portion of the RG  branch as being when a star first exceeds 2.5 times its ZAMS radius. The post-MS evolution is a very sensitive function of the  initial stellar mass. In particular,  the peak radii and timescales of the RG and AGB phases vary considerably with mass. A common feature is that while on the giant branch, stars with a wide range of masses spend considerable time having radii in the range  $\sim 5 - 20 R_\odot$ and brief phases above $100 R_\odot$. The tidal radius of a star, equation \eqref{rt}, is a time-varying function that depends on the mean stellar density, as $\rt(t)\propto M_\ast (t)^{-1/3} R_\ast (t) \propto \bar \rho_\ast (t)^{-1/3}$. The right panel of Figure \ref{fig:EvolveStages} illustrates the variation in stellar evolution profiles with initial mass and that these late  phases of stellar evolution give rise to brief periods in which the star becomes orders of magnitude more vulnerable to tidal disruption than it is on the MS. 

\subsection{Tidal disruption basics applied to evolved stars}\label{sec:SEandTD:imp}

\begin{figure}[tbp]
\begin{center}
\includegraphics[width=0.42\textwidth]{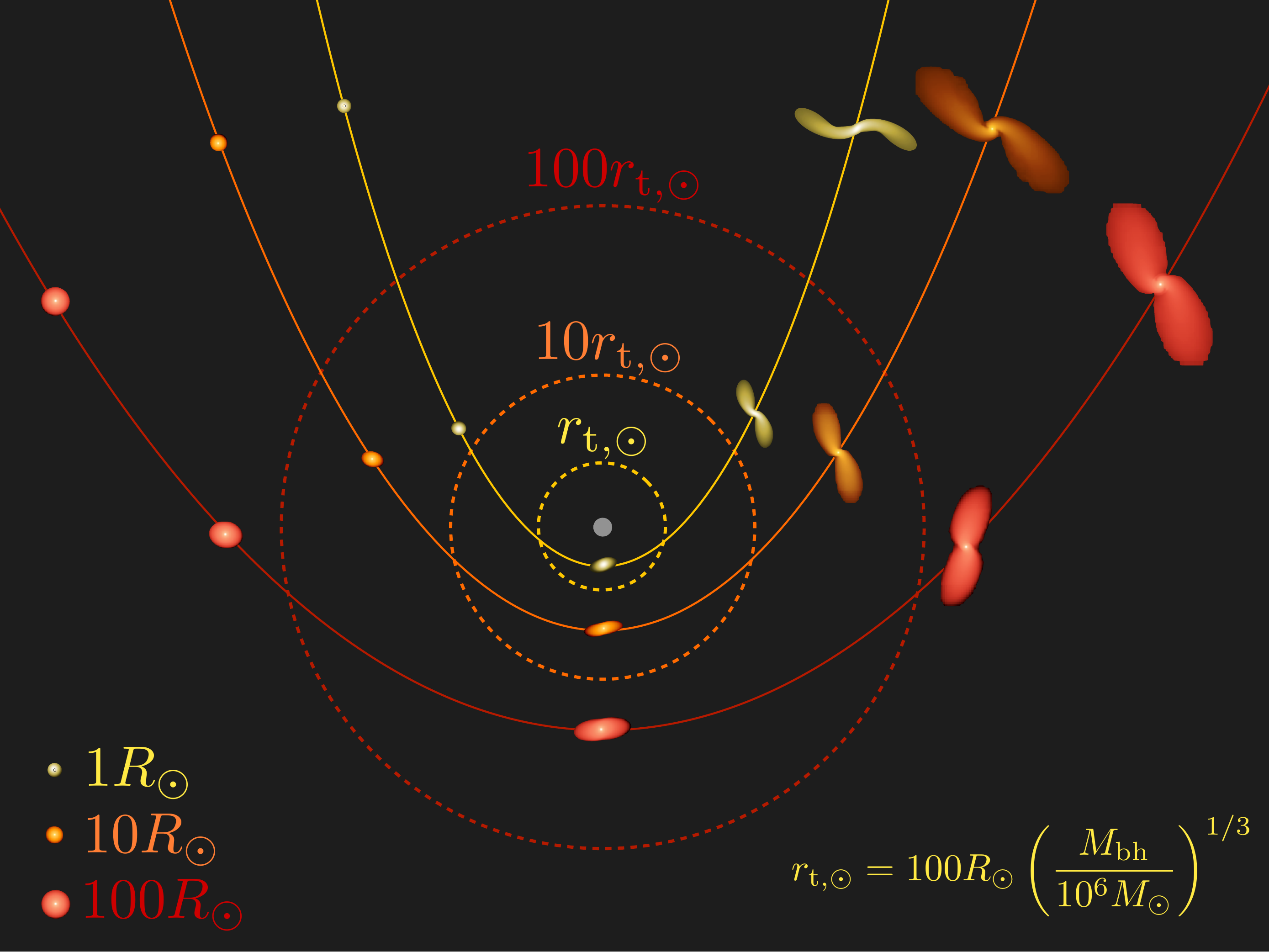}
\caption{A cartoon, {\it not drawn to scale}, highlighting that tidal radius (shown as a dashed line) depends on the mean density of stellar material. As a star evolves off the main sequence and ascends the giant branch its tidal radius (and the corresponding timescales of a tidal encounter) increase by orders of magnitude. Shown here are representative MS (yellow), RG (orange), and AGB (red) stars being disrupted at respectively larger tidal radii.}
\label{fig:RGCartoon}
\end{center}
\end{figure}

Tidal disruptions occur when stars are scattered into sufficiently low angular momentum orbits that they pass within the tidal radius  at pericenter, $\rt = \left( M_{\rm bh} / M_\ast \right)^{1/3} R_\ast$. In Figure \ref{fig:RGCartoon}, we illustrate the respective tidal radii of stars in different evolutionary states.
We denote the impact parameter of an encounter in terms of the ratio of the tidal radius, $\rt$, to the pericenter distance, $\rp$, as $\beta=\rt/\rp$, such that $\beta\gg1$ signifies  a deep encounter. Tidal disruptions are only observable if they occur outside the black hole's Schwarzschild radius $\rs = 2 G M_{\rm bh} / c^2$. In terms of  $\rs$, the tidal radius may be rewritten as
\beq\label{rtrs}
\rt/\rs \approx 23.5 \left(\frac{M_\ast}{M_\odot}\right)^{-1/3}  \left(\frac{M_{\rm bh}}{10^6 M_\odot} \right)^{-2/3} \left(\frac{R_\ast}{R_\odot}\right) .
\eeq
For $M_{\rm bh}\gtrsim 10^8 M_\odot$, solar-type stars will be swallowed whole, producing no observable flare. 
The precise value of this black hole mass cutoff is almost certainly modulated to some extent by the 
black hole's spin and innermost stable circular orbit  \citep{Kesden:2012cn,Haas:2012ci}.
Although stellar remnants such as white dwarfs are expected to be numerous in galactic center environments \citep[e.g.][]{Alexander:2005ij}, we do not consider them here since they will be swallowed whole by black holes $M_{\rm bh}\gtrsim 10^5 M_\odot$ \citep{Luminet:1989uo,Rosswog:2008gc,Rosswog:2008hv}. 

A characteristic encounter timescale in tidal disruption events is the pericenter passage time. This timescale is equivalent to the stellar dynamical time, $\rt/v_{\rm p} \approx \sqrt{R^3/GM} = \tdyn$, for encounters with pericenter at the tidal radius, $\rp = \rt$, regardless of stellar structure or evolutionary state. 
Considering encounters of varying impact parameter, the passage timescale becomes $t_{\rm p} \sim \beta^{-1} \tdyn$.
During the encounter, material is stripped from the stellar core and spread into two tidal tails. The material in one of the tails is unbound from the black hole and ejected on hyperbolic trajectories. The other is bound to the black hole and will return on a wide range of elliptical orbits. Following \citet{Rees1988}, we can derive the fallback time of the most bound material, $\tfb$, by assuming the star is initially on a parabolic orbit, $E_{\rm orb}=0$. Then,  $E_{\rm min} \approx - \beta^2 GM_\ast/R_\ast \left( M_{\rm bh}/M_\ast \right)^{1/3}$,
and the corresponding Keplerian period, $\tfb = 2 \pi G M_{\rm bh} (-2 E_{\rm min} )^{-3/2}$, can be recast as
\beq\label{tfb}
\tfb \approx 0.11 \beta^{-3} \left(\frac{M_{\rm bh}}{10^6 M_\odot} \right)^{1/2} \left(\frac{M_\ast}{M_\odot}\right)^{-1} \left(\frac{R_\ast}{R_\odot}\right)^{3/2} \ {\rm yr}.
\eeq
Accretion flares from tidally disrupted giant stars with $R_\ast \sim10-100R_\odot$ are thus expected to rise and fall on timescales of years to hundreds of years. Since the total mass lost from the star is similar, the peak accretion rate $\dot M \sim \Delta M/\tfb$  for these events will be correspondingly lower than for stars on the MS. 

How often a star enters the zone of vulnerability delineated by the tidal radius depends on the stellar density and velocity dispersion.  One can crudely estimate the rate of tidal disruption as
\begin{align}\label{reesrate}
\dot N_{\rm iso}\sim & 10^{-4} \left(\frac{M_{\rm bh}}{10^6 M_\odot} \right)^{4/3} \left(\frac{n_\ast }{10^5 {\rm \ pc}^{-3}} \right) \nonumber  \\
& \ \ \times \left(\frac{\sigma}{100 \rm{ \ km \ s}^{-1}} \right)^{-1} \left(\frac{\rt}{100R_{\odot}} \right) \ {\rm yr}^{-1},
\end{align}
by assuming that the black hole inhabits an isotropic sea of stars such that $\dot N_{\rm iso} \sim n_\ast \Sigma_{\rm t} \sigma$, and making use of the gravitational focus limit of the tidal disruption cross-section $\Sigma_{\rm t}$ \citep[see][equation 2]{Rees1988}. In equation \eqref{reesrate}, the tidal disruption rate is linearly proportional to the tidal radius, $\dot N_{\rm iso} \propto \rt$. A star that spends most of its lifetime at $1R_\odot$, 10\% of its lifetime at $10R_\odot$, and 1\% at $100R_\odot$ would therefore have a similar likelihood of tidal disruption during each of these stages, even though their timescales  span two orders of magnitude. However, as  noted by \citet{Rees1988}, there are factors which complicate this picture.
In particular, the isotropic description is invalidated by the facts that the number density of stars falls of sharply with distance from the black hole and that stars within the black hole's sphere of influence orbit the black hole directly (in nearly Keplerian orbits). More detailed models  find similar  tidal disruption rates but different scalings than those derived under the simplifying isotropic hypothesis  \citep{Magorrian:1999fd,Wang:2004jy}. The implications of these more accurate models will be discussed in more detail in Section \ref{sec:lc}.

%
%

\section{Hydrodynamics of giant disruption}\label{sec:hydro}

The hydrodynamics of tidal disruption have been studied in detail for  main sequence stars and planets, which are thought to be well described by polytropes.  Initial simulation efforts exploited semi-analytic formalisms that follow the distortion of a sphere into a triaxial ellipsoid in the tidal field of the black hole \citep{Carter:1982fn,Luminet:1985wz,Luminet:1986ch,Kosovichev:1992ud,Ivanov:2001fv}. Later studies used both Lagrangian \citep{Nolthenius:1982dn,Bicknell:1983dn,Evans:1989ju,Laguna:1993cf,Kobayashi:2004kq, Brassart:2008be,Rosswog:2009gg,RamirezRuiz:2009gw,Lodato:2009ib,Antonini:2011ia} 
and Eulerian formalisms \citep{Khokhlov:1993cu,Khokhlov:1993bj,Diener:1997kw,Guillochon:2009di,Guillochon2011,Guillochon2012}. In this section, we will emphasize some of the characteristics that make the disruption of giant stars unique.

\subsection{Initial models for hydrodynamic simulations}\label{sec:hydro:initmodels}

\begin{figure*}[tbp]
\begin{center}
\includegraphics[width=\textwidth]{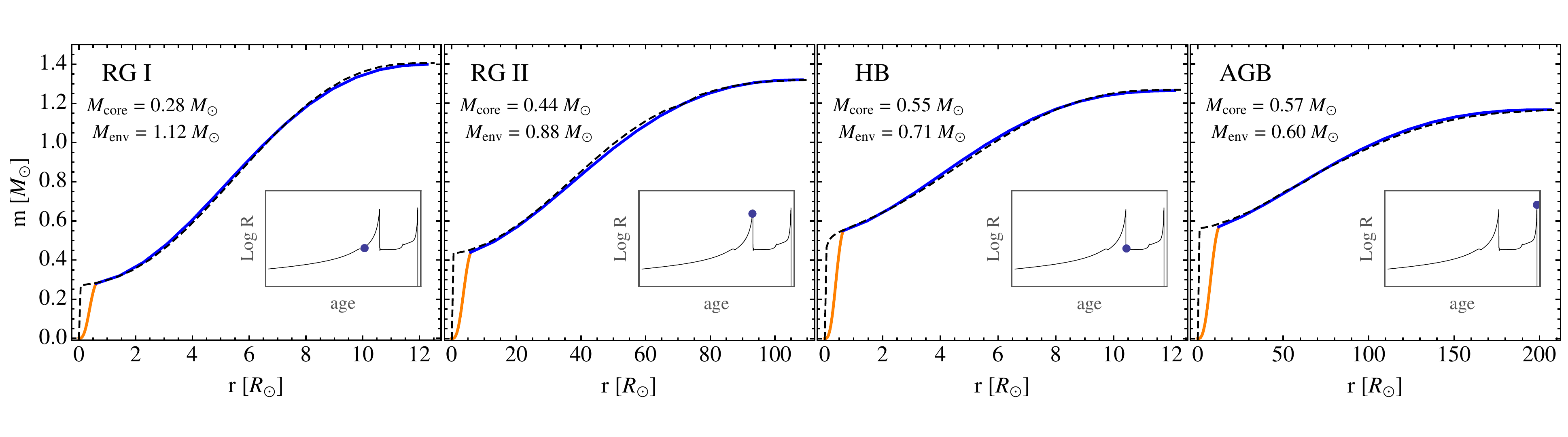}
\caption{In dashed black, we show profiles of enclosed mass calculated in MESA for a $1.4M_\odot$ ZAMS mass star at characteristic stages along its post-MS evolution. From left to right, these are: ascending the RG branch (RG I), the tip of the RG branch (RG II), HB, and AGB. The orange and blue lines show how we approximate the stellar structure using a nested, two-fluid polytrope. We match the profile of the envelope as accurately  as possible while allowing the core to be enlarged $R_{\rm core} \sim R_\ast/20$. Unless the passage is so close as to be disruptive for the core $(\beta>20)$, this approximation of the stellar structure is valid and lessens the computational burden of our calculations.}
\label{fig:mesamatch}
\end{center}
\end{figure*}

We explore the typical characteristics of tidal disruption events involving a $10^6M_\odot$ black hole and stars in the representative stages  outlined in Section \ref{sec:SEandTD} and shown in Figure \ref{fig:EvolveStages}. To this end, we define 4 initial models based on the MESA evolution of  a solar metallicity $1.4 M_\odot$ ZAMS mass star. 
We include two models for RG stars, one of which is ascending the RG branch (model RG I), and one at the tip of the RG branch (model RG II). A single model is used for each of the HB (model HB) and AGB (model AGB) phases of stellar evolution.

For our hydrodynamic calculations we approximate the core and envelope structure using a nested polytropic structure with two fluids of different mean molecular weights \citep{Chandrasekhar:1967vw}. This structure is approximate, but, as seen in Figure \ref{fig:mesamatch}, the fits to the enclosed mass profiles within the envelopes are quite good.  We decided  to enlarge the core radius relative to its true size for numerical convenience because it would be very restrictive to attempt to simultaneously resolve the physical and temporal scales of the true core ($R_{\rm core}\approx 10^9$ cm and $t_{\rm dyn}\approx 5$ s) while also following the evolution of the extended envelope  ($R_{\rm env}\approx 10^{13}$ cm and $t_{\rm dyn} \approx 10^6$ s). In  our models we choose   the core radius to be $R_{\rm core}\approx R_{\rm env}/20$.  This approximation imposes the restriction that we may consider only encounters for which the pericenter distance, $\rp$, satisfies $\rt (R_{\rm core}, M_{\rm core}) \ll \rp \lesssim \rt (R_{\rm env}, M_{\rm env})$. These are encounters that are disruptive for the envelope while leaving the enlarged core tidally unaltered. In Table 1, we give the characteristics of our polytropic approximations to MESA models. 

\begin{deluxetable}{lccccccc}


\tabletypesize{\footnotesize}


\tablecaption{$1.4M_\odot$ ZAMS initial models generated from MESA}

\tablenum{1}

\tablehead{\colhead{Model} & \colhead{$M_\ast$} & \colhead{$M_{\rm core}$ } & \colhead{$R_\ast$ } & \colhead{$n_{\rm core}$} & \colhead{$n_{\rm env}$} & \colhead{$\mu_{\rm   core}/\mu_{\rm env}$} & \colhead{$R_{\rm core}/R_\ast$} \\ 
\colhead{(1)} & \colhead{(2)} & \colhead{(3)} & \colhead{(4)} & \colhead{(5)} & \colhead{(6)} & \colhead{(7)} & \colhead{(8)} } 

\startdata
RGI & 1.4 & 0.28 & 12.5 & 1.5 & 1.7 & 8.0 & 0.048 \\
RGII & 1.32 & 0.44 & 110 & 1.5 & 1.8 & 7.0 & 0.053 \\
HB & 1.26 & 0.55 & 12.3 & 1.5 & 1.7 & 8.0 & 0.052 \\
AGB & 1.17 & 0.57 & 208 & 1.5 & 1.8 & 7.0 & 0.059 
\enddata


\tablecomments{Columns: (1) model name, (2) total mass $[M_\odot]$, (3) core mass $[M_\odot]$, (4) radius $[R_\odot]$, (5) polytropic index of core, where $\Gamma = 1+ 1/n$, (6) polytropic index of envelope, (7) ratio of core to envelope mean molecular weights, (8) ratio of core radius to total radius.}


\end{deluxetable}


\subsection{Methods}\label{sec:hydro:methods}
The calculations  presented here have been performed with the tidal disruption code described in detail in \citet{Guillochon:2009di}, \citet{Guillochon2011} and \citet{Guillochon2012}. Our formalism is based on the FLASH4 code \citep{Fryxell:2000em}, an adaptive mesh Eulerian hydrodynamics code. The compressibility of the gas is  described with a gamma-law equation of state $P \propto \rho^\gamma$. For the envelope we take $\gamma = \gamma_{\rm ad} = 5/3$, while we choose a stiff equation of state to describe the artificial core, $\gamma = 5$. Because the core is only weakly compressible, it does not expand significantly as mass is removed from the star's envelope, ensuring that it makes no spurious impact on the dynamics of the tidal disruption. The flux between grid cells is solved using piecewise parabolic method (PPM) \citep{Colella:1984cg,Woodward:1984ba}, which incorporates second-order directional operator splitting \citep{Strang:1968dr}. We base the grid mesh refinement on the value of the density, with successive cutoffs limiting the refinement by one level per decade in density. 

Our calculations are performed in the frame of the star to avoid introducing artificial diffusivity (for a fixed resolution) by moving the star rapidly across the grid structure \citep{Guillochon:2009di,Guillochon2011}. The self-gravity of the star is computed using  a multipole expansion about the center of mass of the star with $l_{\rm max} =10$. The orbit is then evolved based on the center of mass of the star and the position of the (Newtonian) point mass black hole. We refer the reader to Appendix A of \citet{Guillochon2011} for details of this algorithm. In our simulations the approximation of Newtonian gravity for the black hole is justified by the fact that the closest approach which any of the giant stars in our simulations make is about $110 \rs$, well into the weak field regime. 

The simulations presented here are all resolved by at least $R_\ast/\Delta r_{\rm min}  > 200$, where $\Delta r_{\rm min}$ is the dimension of the smallest cells. However, by implementing density cutoffs in the peak refinement for a given block we allow the linear resolution on the limb of the star to be 2 or 3 levels lower than the peak resolution of the core region. This limits our total  discretization of the computational domain to $(1-10) \times 10^6$ cells, constraining the computational intensity of the problem while allowing us to survey some degree of impact parameter and stellar evolution parameter space. 

\begin{figure*}[tbp]
\begin{center}
\includegraphics[width=0.99\textwidth]{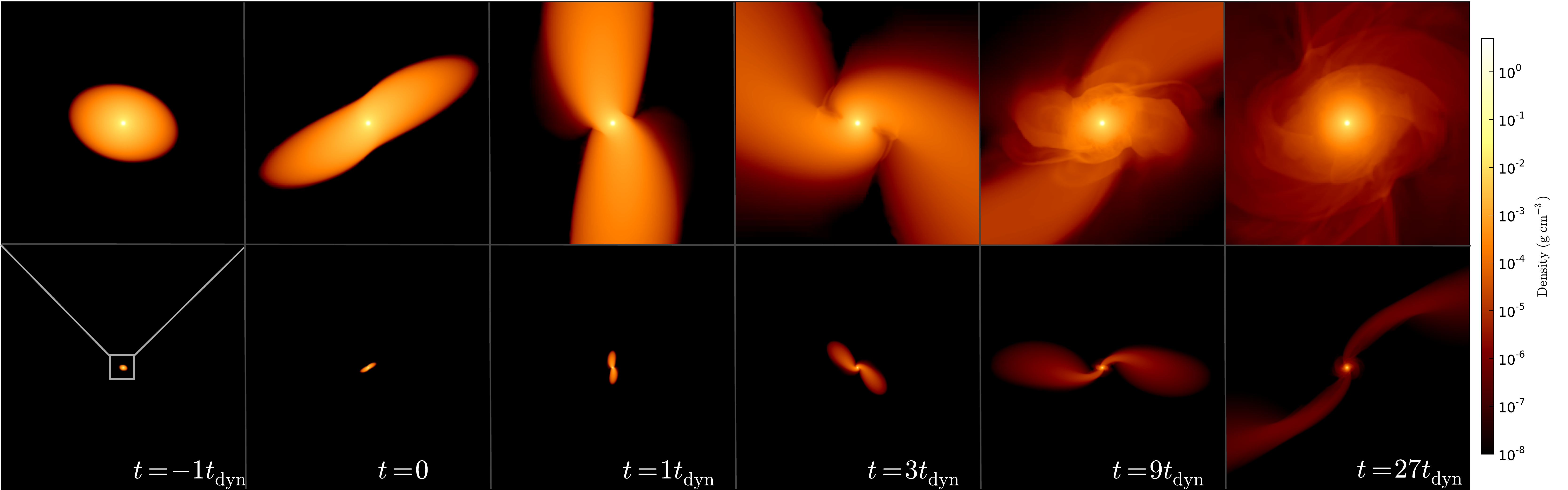}
\caption{Simulation snapshots during a passage through pericenter, t=0, for a $\beta=1.5$ encounter between the RG I model ($1.4M_\odot$,$12.5R_\odot$) and a $10^6 M_\odot$ black hole. The top panels show an inset of the inner region shown in the lower panels, which have widths of 6 and 80 times the initial stellar radius, respectively. }
\label{fig:snapshots}
\end{center}
\end{figure*}

\subsection{Pericenter passage and mass removal}\label{sec:hydro:peri}

The passage through pericenter is mainly characterized by the impact parameter $\beta = \rt/\rp$. Since stars are generally scattered to the black hole from apocenter distances around the SMBH's sphere of influence (see Section \ref{sec:lc}), typical orbits leading to disruption follow nearly radial trajectories. Thus, we consider orbits which are initially parabolic.

As the star passes through pericenter, quadrupole distortions of the stellar surface reach an amplitude of order unity. Material is sheared from the stellar envelope as the time varying tidal field applies a gravitational torque to the distorted star, spinning stellar material to nearly its corotational angular velocity. A portion of this spun-up material is unbound from the star and ejected into two tidal tails. The star's center of mass continues to follow a roughly parabolic trajectory, material in one of the tails is unbound and thus ejected onto hyperbolic orbits, and material in the other tail traces out elliptical orbits and will eventually fall back to the black hole. The tidal field of a $10^6 M_\odot$ black hole is only $\approx 3$\% asymmetric on the scale of a star at the tidal radius (the degree of asymmetry scales linearly with $\beta$). This symmetry is reflected in the formation of  the two similarly shaped tidal tails as shown  in   Figure \ref{fig:snapshots}.

\begin{figure*}[tbp]
\begin{center}
\includegraphics[width=0.99\textwidth]{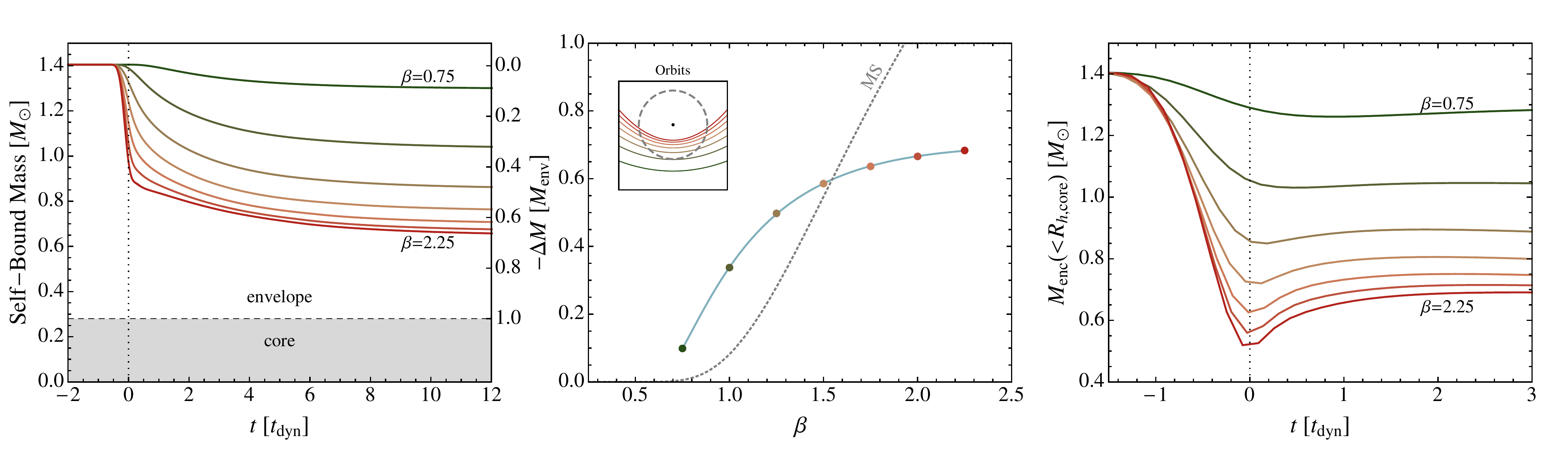}
\caption{Mass loss with varying impact parameter, $\beta$, for the RG I model.  The left panel shows the star's self-bound mass as a function of time, where $t=0$ corresponds to pericenter. The center panel shows the net mass loss, as measured after $\sim 50 \tdyn$, as a fraction of the envelope mass, $M_{\rm env}$.  As $\beta$ increases, the mass lost asymptotes to a value less than $M_{\rm env}$. This can be contrasted to MS stars, which are fully disrupted at $\beta \sim 1.9$. The reason for this asymptotic behavior can be seen in the adiabatic response of the stellar envelope in the expanding background of the core-black hole potential. In the right panel, we plot the enclosed mass within the (time-varying) Hill radius of the core, equation \eqref{rhc}. The increase in enclosed mass after pericenter shows that the envelope is contracting, becoming more dense, and effectively shielding itself from further mass loss. }
\label{fig:DM_RGI}
\end{center}
\end{figure*}

While the structures and tidal deformations seen in  Figure \ref{fig:snapshots} appear broadly consistent with those resulting from the  disruption of MS stars \citep{Guillochon2012}, the quantitative nature of the distortions in the giant star case  is marked by the gravitational influence of the dense core. Of course, MS stars  have cores that are more dense than their envelopes, but the ratio of their central to average  density is  $\rho_{\rm core}/\bar \rho_\ast \approx 100$, whereas for giant stars this ratio is   $\gtrsim 10^6$. In the  initial models we construct, the density contrast is not as high due to numerical limitations, $\rho_{\rm core}/\bar \rho_\ast \approx 2\times 10^3$, but it is similarly marked by a discontinuous transition in the density profile. Because the dynamical time of the core is much less than that of the envelope, the core material is not perturbed, but its gravitational influence significantly modifies the rearrangement of the surrounding  envelope material as the encounter progresses.  The adiabatic response of centrally condensed, nested polytropes, like the ones constructed here, is to contract when they lose sufficient mass, thus shielding the star from further mass loss \citep{Hjellming:1987ci}. The ability of the stellar core to retain  envelope gas  can be seen in the upper panels  in Figure \ref{fig:snapshots}. These panels show a close view of the region near the star's core that leaves the passage tidally excited, but undisrupted. The core itself is completely unperturbed.

In Figure \ref{fig:DM_RGI}, we explore the causes of the mass loss  in greater quantitative detail. The left panel of Figure \ref{fig:DM_RGI} displays the star's self-bound mass as a function of time for a set of encounters with $0.75 <\beta<2.25$.  Encounters of varying $\beta$ lead to different degrees of mass loss and varying speed of mass removal (characterized by the slope of the lines near pericenter, $t=0$). The self-bound mass gradually decreases over many dynamical times, eventually converging to a net mass loss, $-\Delta M$, plotted as a fraction of the envelope mass $M_{\rm env}$ in the center panel of Figure \ref{fig:DM_RGI}. The mass lost during the encounter asymptotes to less than the total envelope mass, with deeper encounters not being able to remove significantly more mass. 

The reason for the observed  asymptotic behavior of $-\Delta M$ with $\beta$ can be inferred from the right panel of Figure \ref{fig:DM_RGI}. The tidal force at pericenter increases with increasing $\beta$, removing a greater fraction of the envelope mass as $\beta$ increases. This increase in the rate of mass loss  is drastically halted when the total envelope mass remaining approaches the mass of the core.  Because the core is a significant fraction of the total stellar mass, the star's structure responds to a drastic loss of  envelope mass by contracting. As more mass is removed, the surrounding envelope contracts more dramatically \citep{Hjellming:1987ci,Passy:2011uc}. The competition between the tidal mass stripping and the ensuing envelope contraction, both of which are seen to increase with $\beta$,  leads to the asymptotic behavior with $\beta$ of $-\Delta M$. A subtlety lies in the fact that this contraction occurs not in an absolute sense, but in an expanding frame defined by the fluid  trajectories in the combined potential of the black hole and the surviving core. This can be seen directly by measuring the  enclosed mass within the Hill radius of the core, shown in the right panel of Figure \ref{fig:DM_RGI}. The Hill radius of the core varies as function of time through the encounter as
\beq\label{rhc}
R_{\rm h,core}(t) = \left(\frac{M_\ast}{M_{\rm bh}}\right)^{1/3} r(t),
\eeq
where $r(t)$ is the separation between the core and the black hole. $M_{\rm enc} (<R_{\rm h,core})$ decreases initially  and reaches a minimum at  pericenter ($t=0$) where $R_{\rm h,core} = \beta^{-1} (M_{\rm core}/M_\ast)^{1/3} R_\ast $. After pericenter, the mass within this equipotential surface increases  because  envelope material contracts relative to the expanding frame of the disrupted star. 

\begin{figure}[tbp]
\begin{center}
\includegraphics[width=0.4\textwidth]{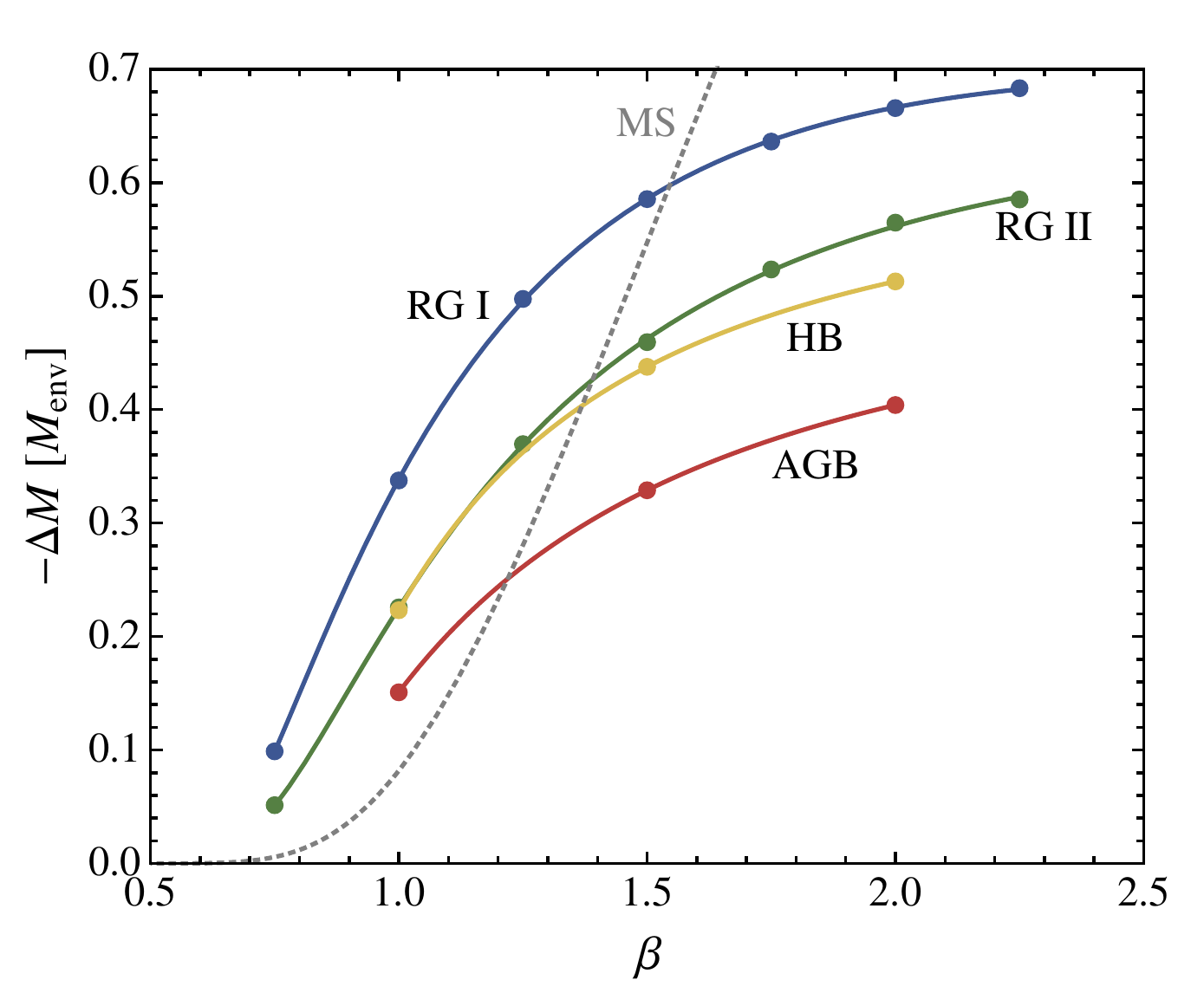}
\caption{Mass loss as a function of $\beta$ for the four post-MS models shown in Figure \ref{fig:mesamatch}. Mass lost is shown as a fraction of the original envelope mass of a given star. Increasing core to envelope mass ratio dictates that that a smaller fraction of envelope mass is lost at a given impact parameter. However, in all cases, much less than the entirety of the envelope mass is lost, even in encounters for which $\beta>1$. The solid lines are described by the fitting formulae presented in equation \eqref{dmfit}. }
\label{fig:DM_all}
\end{center}
\end{figure}

Each of the post-MS models described in Section \ref{sec:hydro:initmodels} differs in the exact polytropic index of the envelope and the core to envelope mass ratio. However, the qualitative nature of mass removal from these models is very similar to that shown in detail for the RG I model in  Figure \ref{fig:DM_RGI}. As can be seen in Figure \ref{fig:DM_all}, evolved stars retain an increasing fraction of envelope material at a given $\beta$ as their core to envelope mass ratio grows. 
This is because larger core mass fractions in evolved stars increase the strength of the envelope's adiabatic response to contract with increasing mass loss \citep{Hjellming:1987ci,Passy:2011uc}. In all cases, a significant fraction of envelope mass remains bound to the core for even the deepest encounters we have explored. 
This stands in stark contrast to MS stars \citep{Guillochon2012}, which are fully disrupted beyond $\beta \sim 1.9$. We tested both more and less dense artificial cores and confirmed that the asymptotic behavior of the mass loss is in no way affected by our enlarged core approximation at the impact parameters we consider.

The following fitting formulae provide a reasonable approximation to the  degree of mass loss within the range of $\beta$ presented here
\beq\label{dmfit}
-\Delta M  = 
\begin{cases}
   0.52 + 1.11\beta^{-1} - 1.95\beta^{-2} + 0.66\beta^{-3}    & \text{RG I, } \\
   0.57 + 0.61\beta^{-1} - 1.58\beta^{-2} + 0.62\beta^{-3}    & \text{RG II, } \\
   0.62 - 0.06\beta^{-1} - 0.30\beta^{-2} - 0.04\beta^{-3}   & \text{HB, } \\
   0.55 - 0.15\beta^{-1} - 0.32\beta^{-2} + 0.07\beta^{-3}    & \text{AGB. } 
  \end{cases}
\eeq

Our findings may be compared to those of authors who study giant star stellar collisions \citep[e.g.][]{Bailey:1999bs,Dale:2009ih}. These studies similarly find that the dense core of the giant star retains a significant fraction of envelope mass. \citet{Dale:2009ih} also note that the great majority of the star's hydrogen envelope must be removed to appreciably alter the star's evolution along the giant branch. Because they retain sufficient portions of their envelope mass, the remnants of the tidal disruption events discussed here do not appear as naked helium cores and instead are likely more similar in structure and appearance to their original, giant-star nature. 

\subsection{Evolution of unbound material}\label{sec:hydro:unbound}

\begin{figure*}[tbp]
\begin{center}
\includegraphics[width=0.99\textwidth]{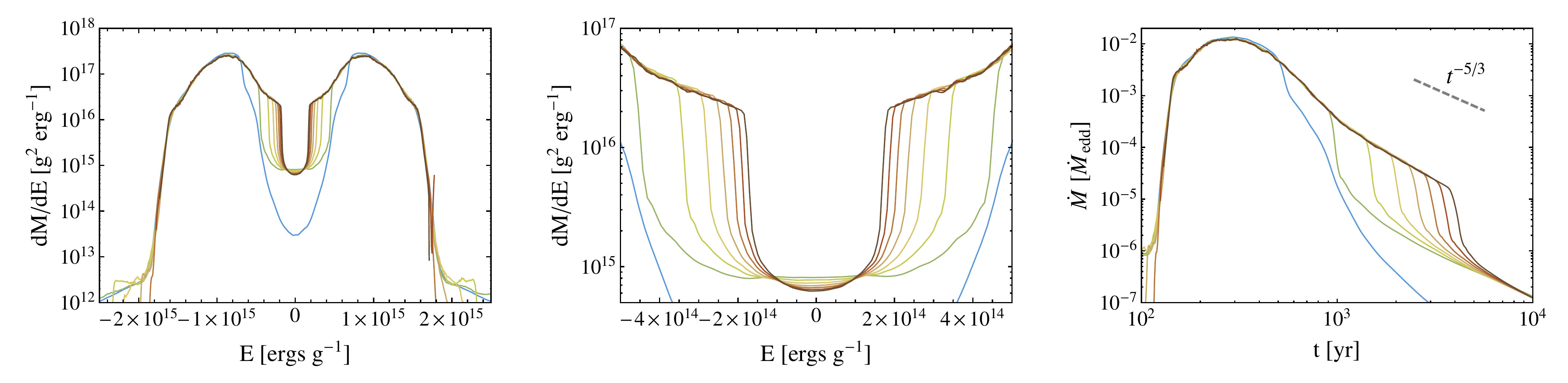}
\caption{Temporal evolution  of the specific binding energy of the tidal tail material, calculated here  for the AGB model at an impact parameter of $\beta=1$. The lines plotted range from $t=5\tdyn$ (blue) after pericenter to $t=35\tdyn$ (brown) after pericenter in intervals of $5\tdyn$ and include only the material which is unbound from the surviving core. In the left and center panels, specific binding energy relative to the black hole is shown on on the x-axis. The star approaches the black hole on a parabolic orbit corresponding to $E=0$. Material with $E>0$ is ejected on hyperbolic orbits, while that with $E<0$ eventually falls back to the black hole on a range of elliptical orbits that may be mapped to the projected fallback rate shown in the right panel using equation \eqref{lodato}. 
The energy distribution of material near the tips of the tidal tails is frozen into homologous expansion at early times. Material closer to $E\sim0$ evolves to fill in the cavity in $dM/dE$ for many $t_{\rm dyn}$ as the surviving remnant becomes increasingly isolated from the black hole's potential. The $dM/dE$ and $\dot M$ distributions shown in Figures \ref{fig:DMDE_time} - \ref{fig:DMDT_all} are plotted with 500 bins in energy or time. }
\label{fig:DMDE_time}
\end{center}
\end{figure*}


Material stripped from the surviving core forms  two symmetric tidal tails (see Figure \ref{fig:snapshots}). The evolution of the  material in the   tails  is well described by two characteristic  stages: an initial period of hydrodynamic evolution, followed by a period of homologous expansion. Hydrodynamic interactions remain important when the local sound speed of the stellar material exceeds the relative expansion velocity of adjacent fluid elements. As the stellar material expands, $\rho$ decreases and the sound speed drops because  $c_{\rm s} = \sqrt{\gamma P/\rho} \propto \rho^{(\gamma -1)/2} \propto \rho^{1/3}$ for $\gamma = 5/3$. When these sound waves are unable to propagate  as quickly as  the distance between neighboring fluid parcels increases, the material's velocity distribution begins to freeze. Any additional evolution of the stellar fluid  is then relatively well described by collisionless trajectories in the black hole's gravitational field. 

To determine the trajectories of the material in the tails, we  calculate the evolution of the specific  binding energies of all of the fluid elements  relative to the black hole. This gives the spread of mass per unit energy within the disrupted star, $dM/dE$, where $E$ is the specific orbital binding energy relative to the black hole. In Figure \ref{fig:DMDE_time}, we plot the time evolution of $dM/dE$ with the self-bound material of the surviving stellar core excised. The majority  of the material achieves homologous expansion only after a few dynamical timescales, with the structure of the material near the tips of the tidal tails being frozen in earliest. 
The distribution of material close to the the surviving core continues to evolve for some time (center panel of Figure \ref{fig:DMDE_time}); it will not fully converge until the surviving remnant is truly isolated from the black hole. This effect can also be seen in the left panel of Figure \ref{fig:DM_RGI}, where the  self-bound mass of the surviving star decreases gradually as a function of time. The gentle evolution of the stream material at the force balance point between the remnant and the black hole occurs  over a long timescale when compared to the pericenter  passage time because the surviving star  continues to be slowly depleted of mass that crosses the lowered potential barrier between the star and the black hole. Hence, the time evolving gravitational influence of the surviving star has to be taken into account to accurately compute the resulting $dM/dE$ distribution.

The material in the tails follows ballistic trajectories after sufficient time has elapsed. As a result,  the spread in energy can be mapped to a return time to pericenter for the tail that is bound to the black hole,
\beq\label{lodato}
\dot M = \frac{dM}{dt} = \frac{dM}{dE}\frac{dE}{dt} = \frac{(2\pi G M_{\rm bh})^{2/3}}{3} \left(\frac{dM}{dE}\right) t^{-5/3}.
\eeq
For a flat  $dM/dE$, one recovers the canonical $t^{-5/3}$ power law predicted for the fallback mass return rate after a tidal disruption event \citep{Rees1988,Lodato:2009ib}. The right panel in Figure \ref{fig:DMDE_time} shows the mapping to $\dot{M}$ from $dM/dE$ plotted  in the left two panels. 
The cavity in $dM/dE$ that results from not including self-bound material leads to a similar cavity  in the predicted fallback rate at very late times. As the remnant flies away from the black hole, becoming increasingly isolated from the black hole's potential, we expect this distribution to fill in (but not necessarily completely flatten).

\subsection{Debris fallback and AGN flaring}\label{sec:hydro:fallback}

\begin{figure}[tbp]
\begin{center}
\includegraphics[width=0.4\textwidth]{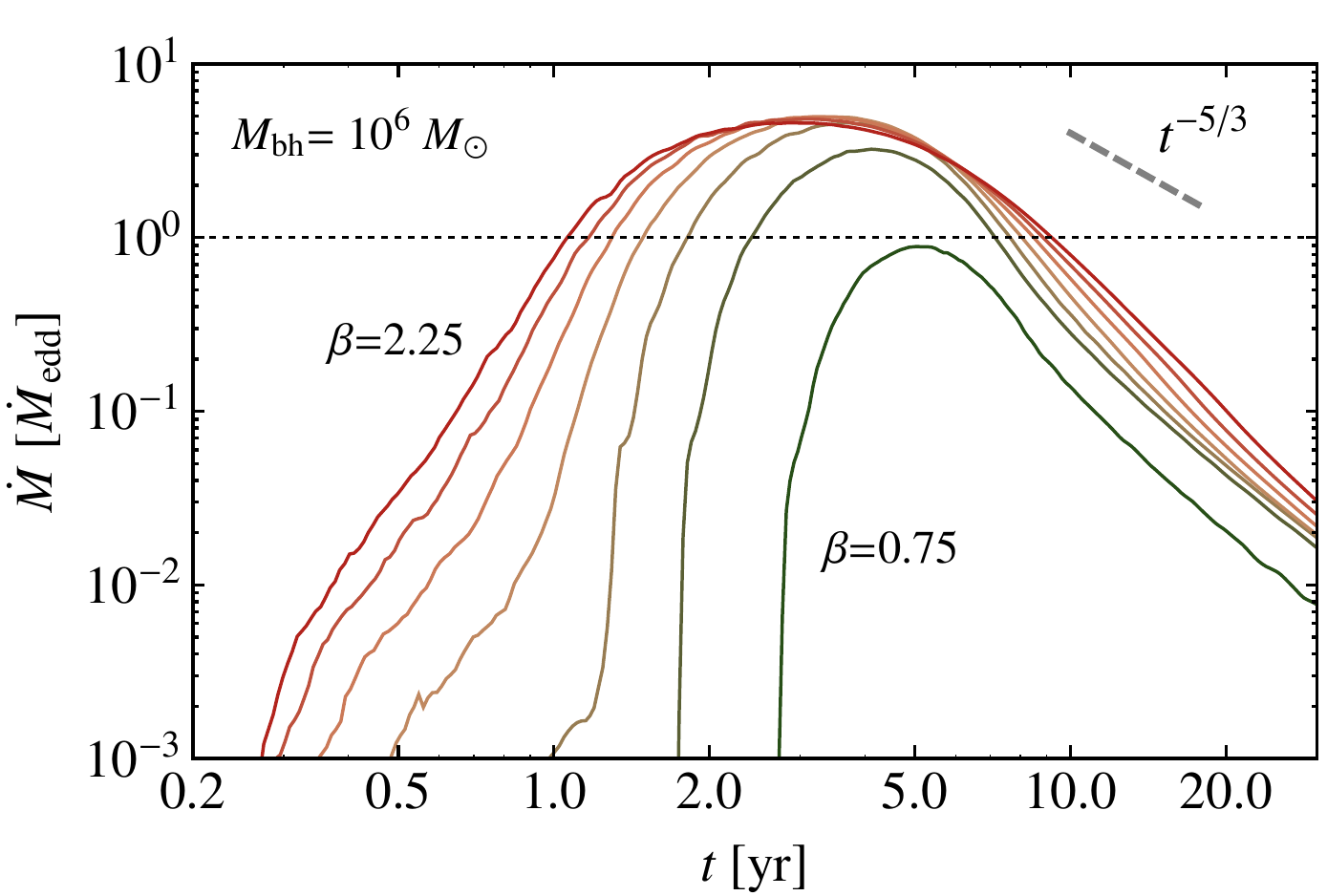}
\caption{The fallback rate of bound material with varying $\beta$ for the RG I model, normalized to the Eddington accretion rate, $\dot M_{\rm edd} = 0.02 \epsilon_{0.1}^{-1} (\Mbh/10^6 M_\odot) \ M_\odot {\rm yr}^{-1}$. This may be compared to Figure \ref{fig:DM_RGI}, which details the removal of mass through pericenter for this model. Differing impact parameter changes $\dot M_{\rm peak}$ and $ t_{\rm peak}$ to some extent. In deeper encounters, the material most bound to the black hole falls back earlier and exhibits a more gradual rise to peak. After peak, all of the $\beta$'s shown here appear to fall off {\it more steeply} than $t^{-5/3}$ for many years. The other three post-MS models exhibit very similar qualitative behavior. }
\label{fig:DMDT_beta}
\end{center}
\end{figure}

\begin{figure}[tbp]
\begin{center}
\includegraphics[width=0.4\textwidth]{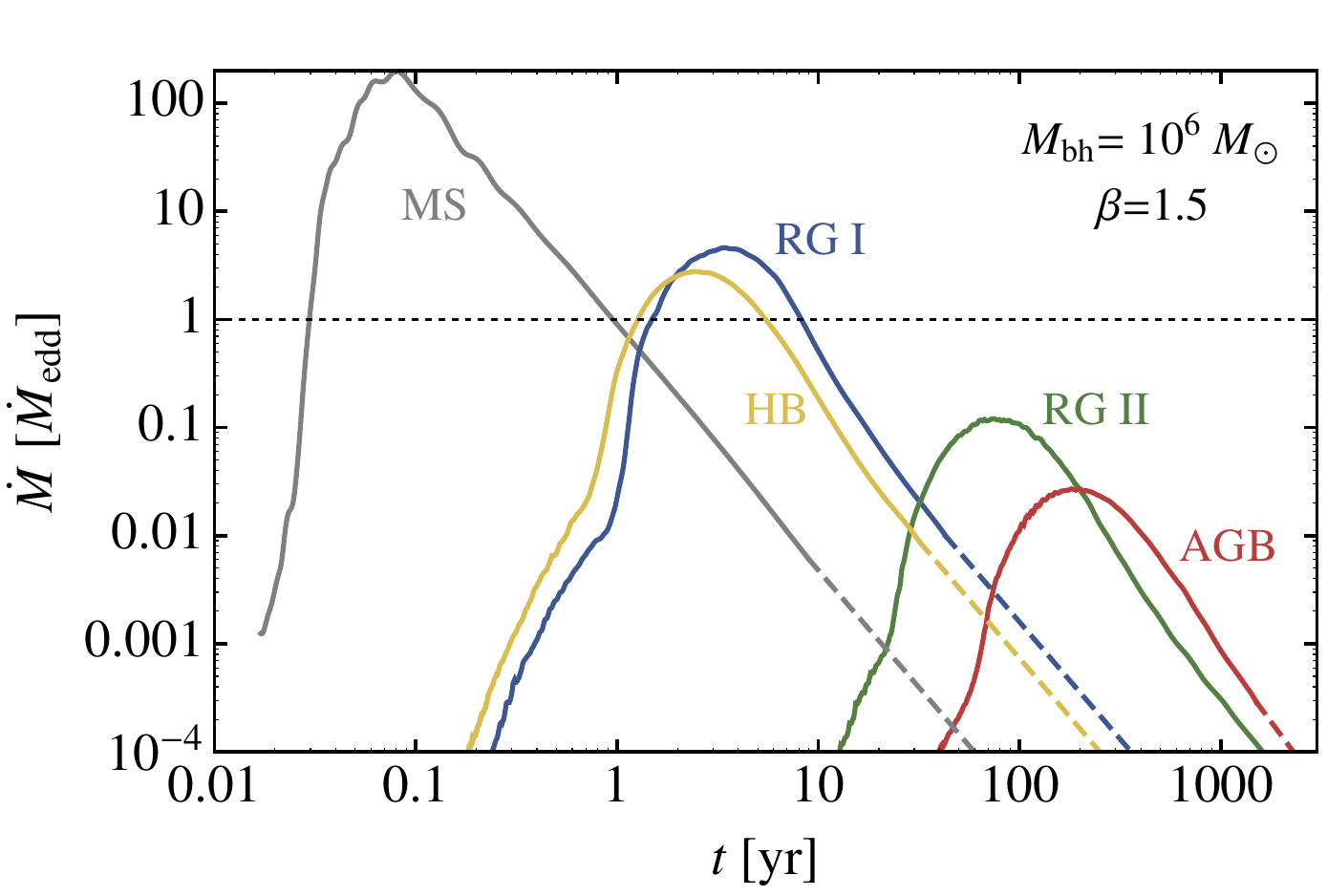}
\caption{A comparison of the fallback accretion between the four evolved star models and the disruption of a sun-like MS star, at $\beta=1.5$. Accretion flares from the disruption of giant stars are long-lived and peak at relatively lower $\dot M$. While the 100-1000 year typical timescales of stars disrupted at the tip of the giant branches will be difficult to distinguish from other long-term low-level AGN activity, The {\it rise to peak} resulting from the disruption of stars either ascending the red giant branch (RG I) or on the horizontal branch stars (HB) should be  observable in current time domain surveys.
We extrapolate the fallback beyond the position of the cavity in $\dot M$ (dashed lines) using the slope of $\dot M$ captured by our last simulation snapshot. 
}
\label{fig:DMDT_all}
\end{center}
\end{figure}

Debris ejected into the tidal tail bound to the black hole returns to pericenter at a rate given by equation \eqref{lodato}.  The returning gas does not immediately produce a flare of activity from the black hole. First material must enter quasi-circular orbits and form an accretion disk  \citep{RamirezRuiz:2009gw}.  Once  formed, the disk will evolve under the influence of viscosity \citep{Cannizzo:1990hw,MontesinosArmijo:2011bl}. However, the viscosity would have to be extremely low  for the bulk of the mass to be stored for longer than $\tfb$ (equation \ref{tfb})  in a reservoir at $r \approx \rt$.   Taking typical values, the ratio of the viscous accretion timescale to the fallback timescale is 
\beq
{t_\nu \over  \tfb} \approx 10^{-3}\beta^{3/2} \left({\alpha_\nu \over 10^{-2}}\right)^{-1}  \left({ \Mbh \over 10^6 M_\odot} \right)^{1/2} \left({M_\ast \over 1M_\odot} \right)^{-1/2}, 
\eeq
where we have made  the standard assumption of a thick $(H/R) \sim 1$ disk \citep{Shakura:1973uy}. The accretion rate onto the black hole, and to a certain extent the bolometric luminosity of the resulting flare (i.e. $L \propto  \dot M c^2$),   is thus generally expected\footnote{When the black hole is fed above the Eddington limit ($\dot{M}>\dot{M}_{\rm edd}$), the relationship between $\dot{M}$ and disk luminosity is, however, less certain \citep[e.g.][]{Strubbe:2009ek,Strubbe:2011iw}.}  to be limited by the gas supply at pericenter $\dot{M}$, equation \eqref{lodato}, not by the rate at which the orbiting debris drains onto the black hole \citep{Rees1988,Ulmer:1999jj}.

In the simulations of giant star tidal disruption presented here, we find sizable variations in the black hole feeding rate, $\dot M$, with impact parameter $\beta$. In Figure \ref{fig:DMDT_beta}, we show $\dot M$  as a function of $\beta$ for the RG I model. The time of peak, $t_{\rm peak}$, is observed to vary only slightly  with $\beta$.  Additionally, since $\Delta M$ asymptotes at high $\beta$, the peak accretion rates, $\dot M_{\rm peak}$, are  similar for most deep encounters. The slope of the initial fallback after peak is, however, significantly steeper than $t^{-5/3}$, particularly for the more grazing encounters. This variation in post-peak slope differentiates $\dot M$ curves that overlap at peak. 

The time of initial fallback, $\tfb$, varies considerably with $\beta$. Because the time of peak, $t_{\rm peak}$, is relatively constant, the slope of the early-time rise varies correspondingly, as can be seen in Figure \ref{fig:DMDT_beta}. Observations of this portion of the $\dot M$ curve could  help break  the relative degeneracy seen in the post-peak light curves. However, at these early times viscous and angular momentum redistribution processes might play some role in shaping the accretion luminosity. 
Further, because the early portion of $\dot M$ represents only a small fraction of the total mass lost by the star, $\Delta M$, it is sensitive in our simulations to the limited mass resolution of discretized material in the tidal tails. In Figure \ref{fig:DMDE_time}, we show that the early-time $\dot M$ freezes into homologous expansion quickly, and is seen to accurately preserve  its shape  as the disruption ensues. We have studied the variations in the early time $\dot M$ with   increasing resolution and find no systematic variations. We do note, however, that the short-timescale bumps and variations  in $\dot{M}$ at $t\sim \tfb$ likely result from grid (de-)refinement  events, which are unavoidable in keeping our computations feasible. 

\begin{figure}[tbp]
\begin{center}
\includegraphics[width=0.35\textwidth]{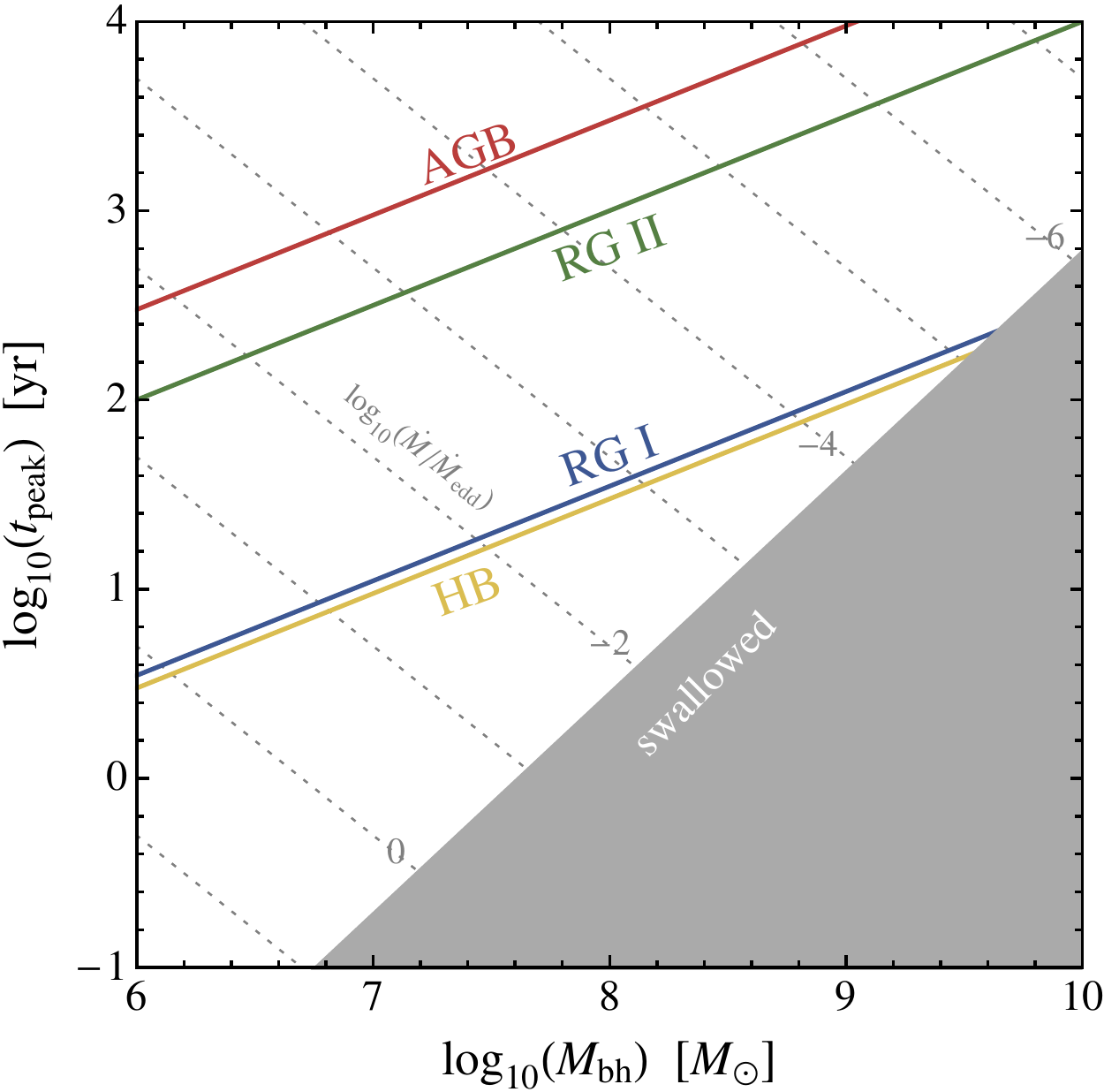}
\caption{ Encounters of $\beta=1$ rescaled to the range of known supermassive black hole masses. With increasing black hole mass, flares become longer-lived but also peak lower relative to the Eddington rate for that black hole mass. The forbidden region shows where $\rt < \rs$ and stars enter the black hole's horizon whole without being tidally disrupted. }
\label{fig:simscalings}
\end{center}
\end{figure}

While variations in $\dot{M}$ with  $\beta$ are present,  major changes  are observed in the feeding rates  as stars evolve and their structures are dramatically altered. In Figure \ref{fig:DMDT_all}, we compare the black hole feeding rates from the disruptions of evolved stars with a typical MS disruption. 
The disruptions of giant stars lead to long-lived flaring events that peak at, or near, the Eddington rate for a $10^6 M_\odot$ black hole. The peak time  of these events is, as expected, correlated with the radial extent of the evolving star. 
Our simulations show that while the timescale of peak accretion rate in  RG I and HB star flares is several years, the rise toward peak $\dot M$ occurs over similar timescales to those seen in the disruption of sun-like  stars, thus offering hope for detectability in surveys tailored to detect MS flares. On the other hand, tidal disruptions of stars at the tips of the giant branches (the RG II and AGB models) lead to flares that peak at hundreds of years timescales, which make them  difficult to  discern from other non-transient AGN feeding mechanisms. In Figure \ref{fig:simscalings}, we show how the peak $\dot M$ and fallback timescales are expected to scale with increasing black hole mass.  

%
%

\section{Loss cone theory and rates of giant disruption}\label{sec:lc}

Following the observation by \citet{Hills:1975kh} that stars passing within $\rt$ at pericenter are tidally disrupted, a series of studies focused on the rate at which stars are fed into disruptive orbits. This rate depends on the distribution of stars around the black hole and the rate at which small scatterings in orbital angular momentum allow stars to diffuse into nearly radial {\it loss cone} orbits that lead to disruption \citep{Frank:1976tg,Lightman:1977hu}. \citet{Cohn:1978cc} present a formalism for treating this diffusion process by numerical integration of a Fokker-Plank equation. More recent work extends the \citet{Cohn:1978cc} formalism to compute the flux of stars into the loss cones of observed galaxies \citep{Magorrian:1998cs,Magorrian:1999fd,Wang:2004jy}, with a key result being that stars are typically fed into the loss cone from weakly bound orbits with semi-major axes similar to the black hole's sphere of influence, equation \eqref{rh}.  \citet{Magorrian:1999fd} also considered the effect of an axisymmetric (rather than spherical) stellar distribution, while \citet{Wang:2004jy} considered the effects of black hole mass on tidal disruption rates. Additional recent work has focused on the effects of black hole spin \citep{Kesden:2012cn}, binary black holes \citep{Ivanov:2005io,Chen:2009bt}, and numerical determination of the tidal disruption rate using n-body calculations \citep{Freitag:2002gw,Brockamp:2011jx}.

\citet{Syer:1999gp} use a simplified treatment of both the loss cone formalism and stellar evolution to roughly account for the effects of evolved stars on the integrated disruption rate. We extend the work of previous authors by emphasizing the relative rates of disruption of stars in different evolutionary phases using our MESA stellar evolution models, but we otherwise follow the Fokker-Plank formalism closely \citep{Cohn:1978cc,Magorrian:1999fd,Wang:2004jy}.
In this section, we begin by describing a simplified galactic nuclear cluster model. We outline the loss cone formalism and compute how the tidal disruption rate changes as a star's tidal radius changes due to stellar evolution. We compare our findings based on the simple nuclear cluster model to a sample of observed nuclear cluster profiles to infer the effects of structural differences between galaxies. We are then able to use our calculation of how the tidal disruption rate scales with tidal radius to determine the stellar ingestion diet of SMBHs. 

\subsection{Simplified nuclear cluster model}\label{sec:KepLC}

To develop intuition for the relative rates of disruption of MS and evolved stars, we first explore a simplified nuclear cluster model. This model consists of a homogeneous stellar population and is described by a Keplerian potential. 
The black hole is the dominant influence on the stellar kinematics within a nuclear cluster of radius
\beq\label{rh}
r_{\rm h} = \frac{G M_{\rm bh}}{\sigma_{\rm h}^2}  = 1.08 \left(\frac{M_{\rm bh}}{10^6 M_\odot }\right)^{1/2.12} \ {\rm pc},
\eeq
where $\sigma_{\rm h}$ is the external velocity dispersion of the greater galactic bulge. In the numerical expression above we use the $\Mbh-\sigma$ relation \citep[e.g. ][]{Ferrarese2000,Gebhardt2000,Tremaine:2002hd,Gultekin:2009hj}, with  fitting values from \citet{Gultekin:2009hj}, 
$\sigma_{\rm h} = 2.43 \times 10^5 (M_\odot/M_{\rm bh})^{4.24}  \ {\rm cm \ s^{-1}}$.

At radii $r \lesssim r_{\rm h}$, the cluster potential is approximately Keplerian,
\beq\label{phi}
\phi(r) = \frac{G M_{\rm bh} }{r} ,
\eeq
where, following \citet{Magorrian:1999fd}, we choose a positive sign convention for the potential and the energies of bound orbits. 
The velocity dispersion at a given radius is approximately $\sigma^2(r) \sim \phi(r)$. We define stellar orbits based on their binding energy to the black hole
\beq
\e(a) = \frac{G M_{\rm bh}}{2 a} = \frac{1}{2}\phi(a) ,
\eeq
where $a$ is the orbital semi-major axis. The Keplerian orbital period of stars with a given $\e$ is
$P(\e) = 2\pi G M_{\rm bh}  (2 \e)^{-3/2} $.

We take a singular isothermal sphere stellar number density profile,
\beq
\nu_\ast(r) = \nu_{\rm h} r^{-2},
\eeq
where $\nu_{\rm h} = \nu_\ast(r_{\rm h})$, as \citet{Wang:2004jy} do. This is comparable to the steep density cusp expected to form in a relaxed stellar distribution about a black hole \citep{Bahcall:1976kk,Bahcall:1977ea}, while also being consistent with the roughly flat velocity dispersion profiles observed in typical galaxies outside $r_{\rm h}$.  For simplicity, a single stellar mass $M_\ast$ is commonly assumed, and, as a result the mass density is related to the number density by $\rho_\ast(r) = M_\ast \nu(r)$. We choose $\nu_{\rm h}$ so that the mass of stars within the sphere of influence is twice the black hole mass, 
which gives $\nu_{\rm h} = (2 \pi r_{\rm h}^3)^{-1}  (M_{\rm bh}/M_\ast) $.
We further assume that stellar orbits are isotropic in angular momentum space and therefore may be described by a distribution function solely in energy,
\beq
f(\e) = (2\pi \sigma_{\rm h}^2)^{-3/2} \frac{\Gamma(3)}{\Gamma(1.5)} \left(\frac{\e}{\sigma_{\rm h}^2}\right)^{1/2} ,
\eeq
where the numerical factor $\Gamma(3)/\Gamma(1.5) \approx 2.26$ \citep[equation 9 in][]{Magorrian:1999fd}. 
Having described the properties and phase space distribution of stars in our cluster model, we can now determine how often stars are disrupted.

\subsubsection{The full loss cone disruption rate}
The angular momentum of a circular orbit is the maximum angular momentum allowed for a given $\e$. It is
\beq
\Jc^2(\e) = \frac{G^2 M_{\rm bh}^2}{2\e} .
\eeq
The angular momentum of a loss-cone orbit is 
\beq
\Jlc^2(\e) = 2 r_{\rm min}^2 \left(\frac{G M_{\rm bh}}{r_{\rm min}} - \e \right) \approx 2 G M_{\rm bh} r_{\rm min},
\eeq
where $r_{\rm min}$ is the maximum of $\rt$ or $\rs$.

The number of stars in a full loss cone per energy is then $N_{\rm lc}(\e) =  4 \pi^2 f(\e) P(\e) \Jlc^2(\e)$, and 
the total number  is
\beq
\mathcal{N}_{\rm lc} = \int_0^{\e_{\rm max}} N_{\rm lc}(\e) d\e, 
\eeq
where $\e_{\rm max}$ is the energy corresponding to $r_{\rm min}$. 
The flux of stars into the loss cone, if we assume that it  is continuously replenished, is simply
\beq
F_{\rm full}(\e) = N_{\rm lc}(\e)/P(\e) .
\eeq
Similarly, the net rate at which stars are disrupted from a full loss cone is obtained by integrating over energy,
\beq
\dot N_{\rm full} = \int_0^{\e_{max}} \frac{N_{\rm lc}(\e)}{P(\e)} d\e. 
\eeq
However, the loss cone is not necessarily full for all $\e$. We turn or attention to where $\dot N_{\rm full}$ is an appropriate estimate of the tidal disruption rate.

\subsubsection{Repopulating loss cone orbits}

The derivation of $\dot N_{\rm full}$ assumes that at all energies, $\e$, stars destroyed by the black hole at pericenter are replaced on a timescale shorter than the orbital period, $P(\e)$, thus ensuring that the phase space density of low $J$ orbits remains undepleted. A typical timescale for stars to random walk in angular momentum by $\Jlc$ via the two-body relaxation process is 
\beq\label{tj}
\tJ(\e) \sim \left[\frac{\Jlc(\e)}{\Jc(\e)}\right]^2 \tr\left(a(\e)\right),
\eeq	
where two-body relaxation time is
\beq
\tr(r) = \frac{5.68 \times 10^{19} }{{\rm ln} \ \Lambda} \frac{M_\odot}{M_\ast} \frac{10^4 M_\odot {\rm \ pc^{-3}}}{\rho_\ast(r)}  \left(\frac{\sigma(r)}{10^7 \rm {cm/s}} \right)^3 \ {\rm s} ,
\eeq
\citep[equation 7-107 in][]{Binney2008}.  
The Coulomb logarithm is approximated as 
$\Lambda \approx 0.4 \mathcal{N}_*(r<r_{\rm h})$, where $ \mathcal{N}_*(r<r_{\rm h})$ is the number of stars within $r_{\rm h}$, or equivalently, the number of stars with $\e > \e_{\rm h}$.

Thus, in an orbital period, the typical change in angular momentum a star receives via this random walk process is
\beq\label{dj}
\Delta J^2 (\e) = \Jlc^2(\e) P(\e)/\tJ(\e),
\eeq
or, written as a ratio to the loss cone angular momentum \citep{Cohn:1978cc},
\beq
q(\e) \equiv \frac{\Delta J^2 (\e)}{\Jlc^2 (\e)} = \frac{ P(\e)}{\tJ(\e)}.
\eeq
We can imagine two limiting cases for the value of $q$. The full loss cone rate is appropriate when $\Delta J \gg \Jlc$. This is the so called pinhole limit where per-orbit scatterings are large compared to the target, here given by the size of the loss cone \citep{Lightman:1977hu}. Loss cone orbits are easily repopulated in an orbital time in this limit, ensuring that $F_{\rm full}$ is satisfied. In the opposite limit, $\Delta J \ll \Jlc$, the loss cone refills much more gradually than it is depleted. Once in the loss cone, stars have little chance of being scattered out in an orbital period. They are disrupted as they pass through pericenter and the loss cone remains mostly empty. This is known as the diffusion limit, where stars must diffuse toward the loss cone over many orbital periods \citep{Lightman:1977hu}. 

The flux into the loss cone in the diffusion limit is less than the full loss cone rate, since the loss cone will  be depleted. We define the parameter $R_0$ to specify the low angular momentum limit above which orbits are typically populated,  
\beq
\frac{R_0(\e)}{R_{\rm lc}(\e) } = 
	\left\{
	\begin{array}{cl}
	{\rm exp}\left(-q\right) & \quad \text{if $q(\e)>1$,} \\
	{\rm exp}\left(-0.186 q -0.824\sqrt{ q }\right) & \quad \text{if $q(\e)<1$,}
	\end{array}  \right.
\eeq
where $R \equiv J^2/\Jc^2$ \citep{Magorrian:1999fd}. This allows us to calculate the loss cone flux in the transition between the two limiting cases, where loss cone orbits are partially repopulated in an orbital period. The flux of stars into the loss cone is given by
\beq\label{Flc}
F_{\rm lc}(\e) = 4\pi^2 \Delta J^2(\e) \frac{f(\e)}{{\rm ln}(R_0^{-1})}
\eeq
\citep{Magorrian:1999fd}. In the pinhole limit, $R_0(\e) \to 0$, and $F_{\rm lc}(\e) \to F_{\rm full}(\e)$. For more tightly bound orbits  corresponding to the diffusion limit we have  $R_0(\e) \to R_{\rm lc}(\e)$ and $F_{\rm lc}(\e) \ll F_{\rm full}(\e)$.
Finally, the integrated flux of stars into the loss cone is obtained by integrating $F_{\rm lc}(\e)$ over $\e$,
\beq\label{Ndot}
\dot N = \int_0^{\e_{\rm max}} F_{\rm lc}(\e) d\e.
\eeq

\begin{figure}[tbp]
\begin{center}
\includegraphics[width=0.4\textwidth]{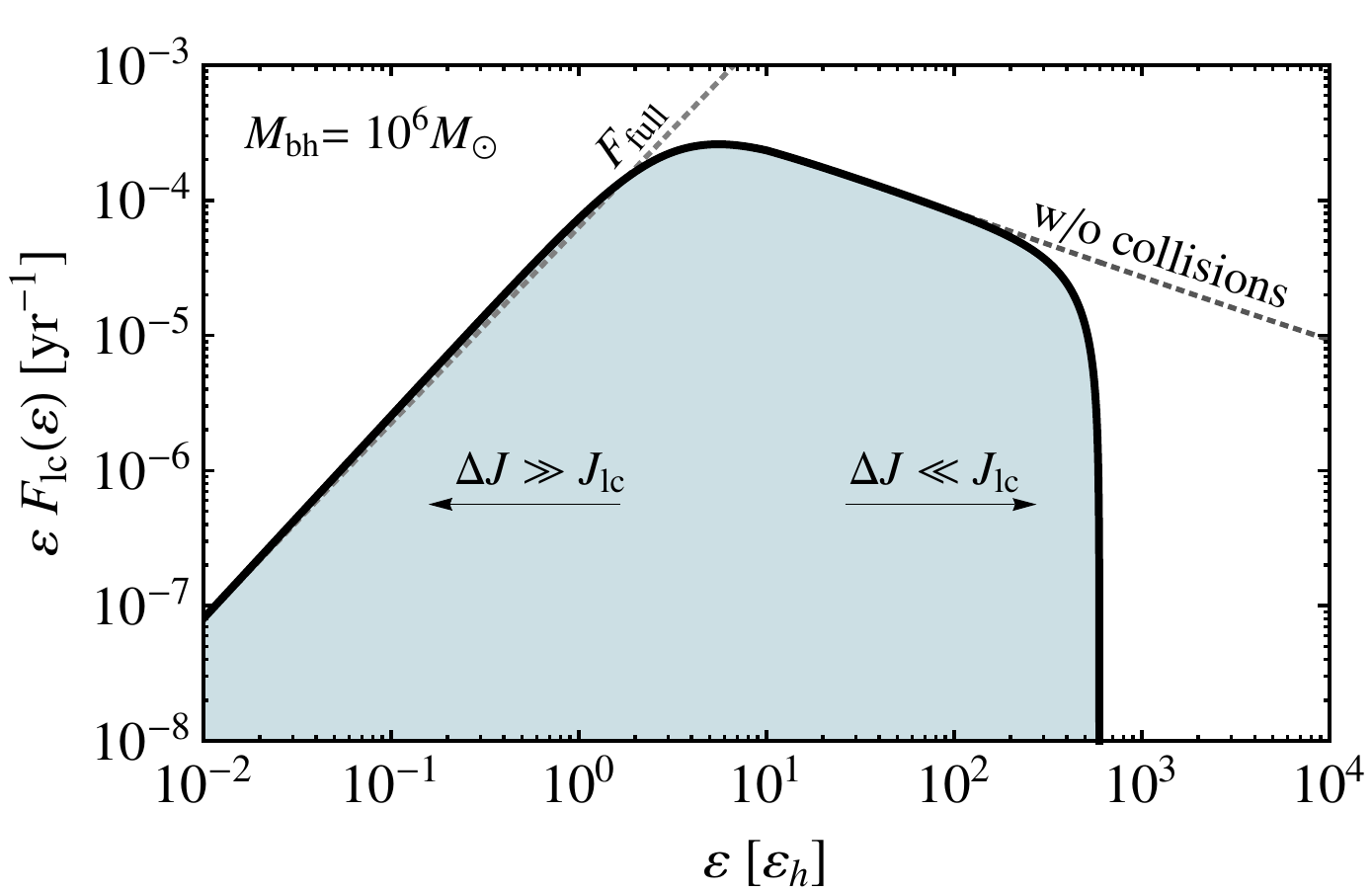}
\caption{Loss cone flux $F_{\rm lc}$ of solar-type stars for the simplified nuclear cluster model of Section \ref{sec:KepLC}. Orbital binding energy is inversely proportional to semi-major axis $\e\propto 1/a$, and $\e_{\rm h}$ is the energy corresponding to the radius of influence of the black hole, in this case $r_{\rm h} \sim 1$ parsec. At low $\e$, $\Delta J\gg \Jlc$, and the loss cone is refilled on an orbital timescale. In this limit, $\Flc$ follows the full loss cone rate. At higher $\e$,  $\Delta J \ll \Jlc$, and orbits with $J<\Jlc$ are incompletely repopulated in an orbital period. In this limit, $\Flc$ is set by the diffusion rate in orbital angular momentum. At $\e \gg \e_{\rm h}$ a star must complete $N_{\rm orb} \sim \Jlc/\Delta J$ orbits to diffuse one loss cone angular momentum, $\Jlc$. At each pericenter passage stars are vulnerable to collisions, leading to a collisional truncation of $F_{\rm lc}$ where $N_{\rm orb} \gg 1$. }
\label{fig:FlcKep}
\end{center}
\end{figure}

Figure \ref{fig:FlcKep} shows the loss cone flux for the simplified nuclear cluster model described here. The peak of the flux comes from $\e \sim \e_{\rm h}$, which corresponds to orbits with semi-major axes comparable to the black hole's sphere of influence, $r_{\rm h}$. At energies lower than the peak, the pinhole limit is applicable, and $\Flc \sim F_{\rm full}$. At high energies, $F\ll F_{\rm full}$ because the loss cone is only partially refilled each orbital period. 

\subsubsection{Modification of $F_{\rm lc}$ by direct stellar collisions}

In the crowded nuclear cluster environment, the large geometric cross-section of giant stars leaves them susceptible to collisions with other stars. The effect of these collisions on $F_{\rm lc}$ is most pronounced in the diffusion limit, where stars must complete $N_{\rm orb}\sim1/q(\e) = \tJ(\e) / P(\e)$ orbits to be able to diffuse a total  $\Delta J=\Jlc$ in angular momentum space. To estimate the importance of collisions, we compute the probability of survival, $P_{\rm surv}$, during one angular momentum diffusion time, $\tJ$. The collision cross section is 
\beq
\Sigma(R_{\rm min},r) = \pi R_{\rm min}^2 \left[ 1 + \frac{2G(M_1 + M_2)}{R_{\rm min} \sigma^2(r)}  \right],
\eeq
where $R_{\rm min} = R_1+R_2$ and the two stars are represented by subscripts 1 and 2.  We approximate low angular momentum  orbits (which may diffuse to loss cone orbits within $\tJ$) as radial paths, $s$, through the cluster from apocenter to pericenter and back again. The typical number of collisions suffered during  $\tJ$ is then
\beq
 N_{\rm coll}(\tJ) = \frac{2}{q(\e)} \int_{\rp}^{2a(\e)} \Sigma(R_1+R_2, s) \nu(s) ds ,
\eeq
and the probability of survival is
$
P_{\rm surv}(\tJ) = {\rm Max}\left[0,1 -\eta N_{\rm coll}(\tJ) \right],
$
where $\eta$ represents the efficiency of collisions in destroying stars. For our purposes, $\eta = 1$ is a reasonable choice because even if stars are not completely destroyed by a collision, they suffer a large angle scattering sufficient to deviate their orbital parameters significantly from their previous random walk. 

The importance of collisions can be seen in Figure \ref{fig:FlcKep} (which considers solar-type stars). At high orbital binding energies, $\e$, $\Flc$ is truncated by the strong likelihood of stars in low $J$ orbits suffering a collision in $\tJ$.

\subsubsection{Scaling of the loss cone flux with tidal radius}\label{sec:scale}

To determine the relative contribution of different stellar evolutionary stages to the tidal disruption rate, we consider the per star (specific) tidal disruption rate $\dot n$. The specific rate, $\dot n$, varies over the stellar lifetime with changing tidal radius, and, because we consider a cluster of homogeneous stellar constituents, $\dot n  = \dot N/N$.  In Section \ref{sec:SEandTD:imp}, we show that for an isotropic  distribution  of stars, the flux  into the loss cone scales linearly with the tidal radius. This is only true because the loss cone is full, thus the rate depends on the size of the loss cone, and $\dot n \propto \Jlc^2 \propto \rt$. If the loss cone is empty, changes in its size do not have an effect on the rate, which is set instead by the diffusion process. In the terminology of \citet{Lightman:1977hu}, these are the pinhole and diffusion limits, respectively.
Taking the relevant limits of equation \eqref{Flc}, we have
\beq\label{FlcLims}
F_{\rm lc}(\e) \propto 
\begin{cases}
\Jlc^2 \propto \rt & \text{for } q(\e) \gg 1\; \text{(pinhole)}, \\
\text{ln}\left( \Jlc^2 \right) \propto \text{ln}\left( \rt \right) & \text{for } q(\e)\ll 1\; \text{(diffusion)}.
\end{cases}
\eeq
The exact scaling the tidal disruption rate, $\dot n$, must be some combination of the contribution from these two limiting cases because it is an integral quantity in $\e$ that spans both limits.
 
\begin{figure*}[tbp]
\begin{center}
\includegraphics[width=0.99\textwidth]{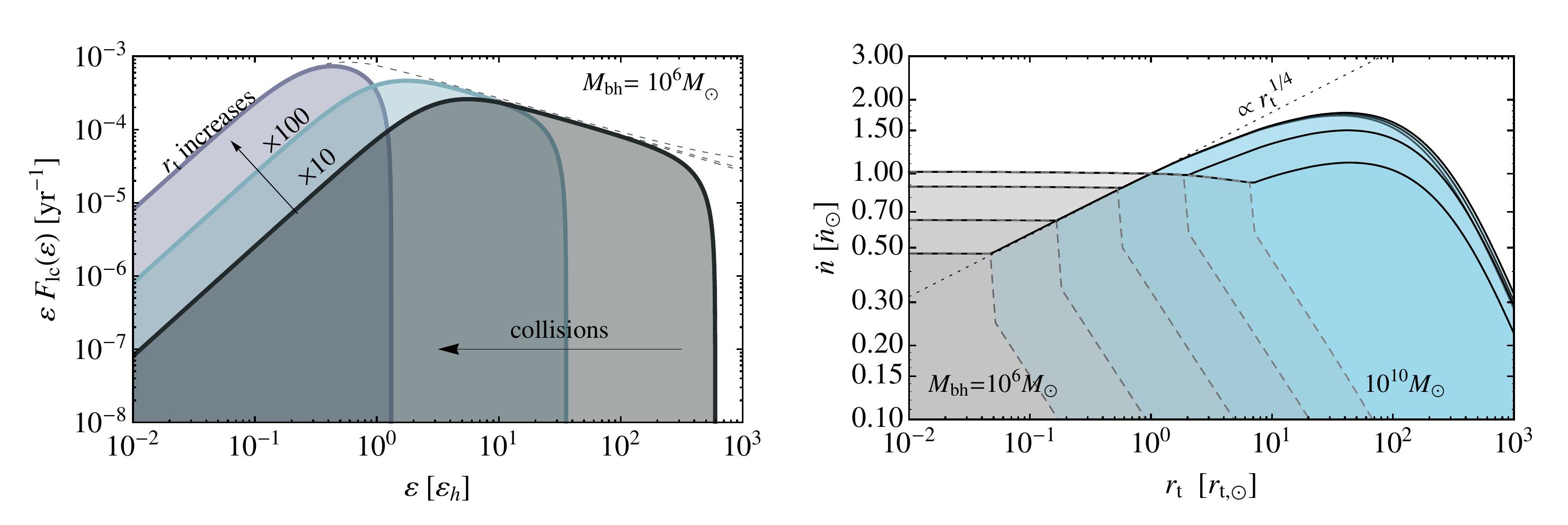}
\caption{Scaling of the tidal disruption rate with increasing tidal radius. The left panel shows that with increasing $\rt$ in the $\Delta J\gg \Jlc$ limit, $F_{\rm lc}$ scales linearly with $\rt$, while in the $\Delta J\ll \Jlc$ limit, $F_{\rm lc}$ scales only weakly with $\rt$. We assume that increasing stellar radius is the mechanism causing $\rt$ to increase  (because in stellar evolution the changes in mass are small). As a result, for larger $\rt$ collisions become increasingly important.  The right panel shows the integrated rate per star, $\dot n$. For a wide range of tidal radii, the scaling is well described by a power law with slope, $\alpha = 1/4$. At very large $\rt$ collisions become dominant and $\dot n$ decreases, while for small $\rt$ stars are typically swallowed whole by the black hole because $\rs > \rt$, and $\dot n$ no longer depends on $\rt$. The shaded regions describe the relative fractions of swallowed (grey) versus flaring (blue) events for a range of black hole masses between $10^6 M_\odot$ and $10^{10} M_\odot$.  }
\label{fig:FlcKepScalingRt}
\end{center}
\end{figure*}

 The effect of the superposition of the full and empty loss cone limits within a particular nuclear cluster is shown in Figure \ref{fig:FlcKepScalingRt}. The left panel shows the change in the loss cone flux, $\Flc$, as $\rt$ increases by a factor of 10 and 100.  In the portion of the cluster where $\Flc \sim F_{\rm full}$, $F_{\rm lc}$ scales linearly with $\rt$. In the empty loss cone limit $F_{\rm lc}$ is approximately constant with increasing $\rt$, as expected. Because $\Jlc$ is increasing while $\Delta J$ remains the same, the energy corresponding to $q(\e)=1$ (roughly the peak in $\Flc$) moves to lower $\e$. Therefore, as the loss cone increases in size, stars are typically fed from less bound orbits that originate further from the black hole. 

Specific tidal disruption rates, $\dot n$, can be calculated via the integrated loss cone flux, $\dot N$, equation \eqref{Ndot}.
This integral is shown for a range of $\rt$ and $M_{\rm bh}$ in the right panel of Figure \ref{fig:FlcKepScalingRt}, normalized to the specific rate of solar-type stars for a given $M_{\rm bh}$. For a wide range of tidal radii, the integrated rate follows $\dot n \propto \rt^\alpha$, with $\alpha = 1/4$. At large black hole masses $r_{\rm s}$ is larger than $r_{\rm t,\odot}$, and, as a result, solar-type stars are swallowed whole rather than producing a disruption, as described by equation \eqref{rtrs}. This is coded in  the right panel of Figure \ref{fig:FlcKepScalingRt} as grey shaded regions for stars consumed by the black hole and blue shaded regions for flaring events.  For stars larger than $\sim 100 R_\odot$ collisions become a dominant effect and $\dot n$ substantially decreases.

\subsection{Impact of the structural diversity of observed galactic centers}\label{sec:RealGal}

In Section \ref{sec:KepLC}, we compute the relative likelihood of tidal disruption for stars of varying tidal radii under the assumption that the stellar density distribution is accurately described by a singular isothermal sphere. However, the central regions of observed galaxies exhibit somewhat more complex structure. Out of this complexity arises some degree of galaxy to galaxy variation, which we illustrate here. 

\subsubsection{Galaxy sample}
Observations resolving the central SMBH's sphere of influence in nearby, early type galaxies have been performed by the Nuker team using the {\it Hubble Space Telescope} (see \citet{Lauer:1995dm} for early work and \citet{Lauer:2007el} for a more recent review). The radial surface brightness profile in these galaxies is found to be bimodal \citep{Faber:1997fn}. Many of the most massive galaxies exhibit a core or flattening of the surface brightness profile at small radii. The presence of such cores is thought to be lingering evidence of gravitational heating from the inspiral of a pair of black holes following a major galactic merger \citep{Faber:1997fn}. The surface brightness profile of less massive galaxies typically continues to rise at smaller radii up to the resolution limit of the observations and is well described by a single power law. These galaxies are termed power-law and exhibit central stellar densities tens to thousands of times higher than those inferred for core galaxies \citep{Lauer:2007el}. 
Nuker galaxy surface brightness profiles are fit to a parameterized, smooth function (the Nuker law) that describes the asymptotic slopes of the surface density profile at large and small radii relative to a break radius $\rb$ \citep{Byun:1996hw}. The rates of tidal disruption events in these galaxies are computed by \citet{Magorrian:1998cs} and \citet{Magorrian:1999fd} and revised by \citet{Wang:2004jy} to accommodate changes in inferred black hole masses.

\begin{figure*}[tbp]
\begin{center}
\includegraphics[width=0.99\textwidth]{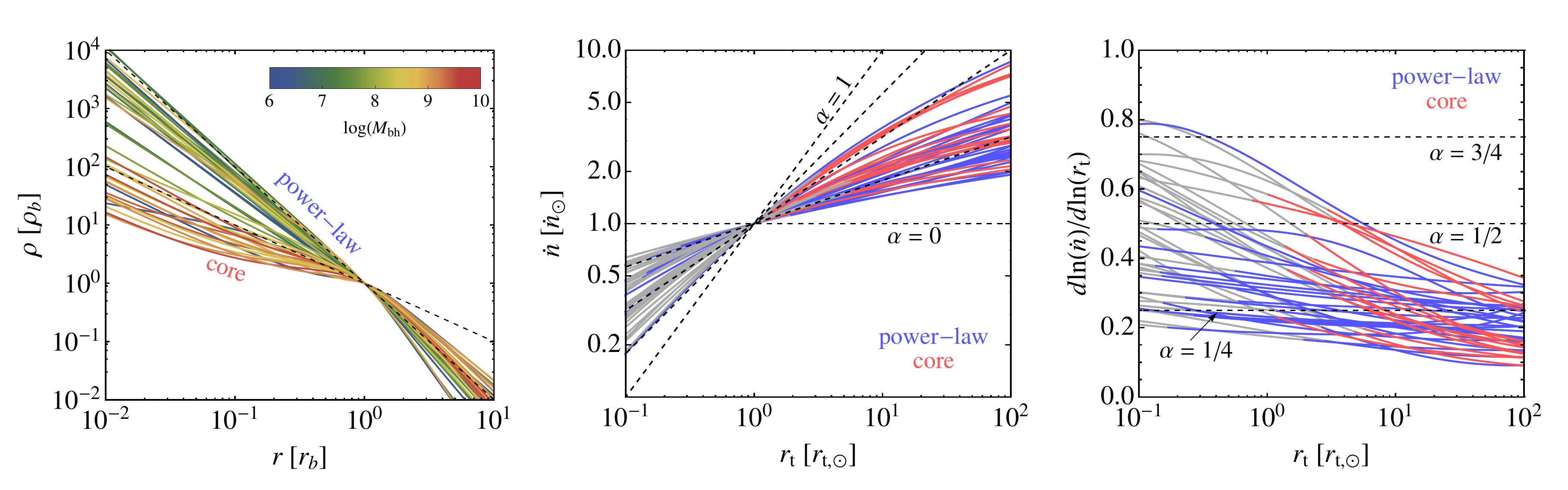}
\caption{Scaling of the tidal disruption rate in a sample of 41 early type galaxies whose central surface density profiles were observed with {\it HST} and fit with parameterized {\it Nuker} laws \citep[for the details of the sample see][]{Wang:2004jy}.
In the left panel, we show the mass density of stars as a function of the density at the break radius, $\rb$, a parameter of the two-power Nuker fit. Inside $\rb$, galaxies show bimodal structures, with those hosting the most massive black holes typically having low density cores, whereas those with less massive black holes typically have dense central regions and single power law profiles. In the center panel, we plot the scaling of the specific tidal disruption rate $\dot n$ with $\rt$ for each of these galaxies, normalized to the rate of solar type stars, $\dot n_\odot$. Where the tidal radius is less than the Schwarzschild radius we still compute the rate for illustrative purposes, but color the lines in gray rather than by galaxy structure. Finally, the right panel shows the slope of the scaling curves plotted in the center panel as a function of $\rt$. There is more substantial galaxy to galaxy variation in the slopes of these scaling curves than there is between structural class of galaxies. However, in most cases, $d {\rm ln} ({\dot n}) / d {\rm ln} (\rt) \ll 1$ and in many cases is broadly consistent with the $\alpha = 1/4$ scaling derived in the simplified model of section \ref{sec:KepLC}. }
\label{fig:NukerGals}
\end{center}
\end{figure*}

\subsubsection{Scaling of $\dot n$ with $\rt$}
To investigate the influence of the bimodal nature of galactic center profiles on the scaling of $\dot n$ with $\rt$, we follow the formalism  of \citet{Magorrian:1999fd} and \citet{Wang:2004jy}. We use the galaxy sample from \citet{Wang:2004jy}, who used data from \citet{Faber:1997fn} and fit black hole masses based on the $\Mbh-\sigma$ relation of \citet{Merritt:2001kj}.
We show the deprojected mass density profiles of the galaxy sample used in our analysis in the left panel of Figure \ref{fig:NukerGals}. At radii less that the break radius, $\rb$, the mass density profiles of core and power-law galaxies separate dramatically. Black holes with the highest inferred masses are typically found in core galaxies, rather than power-law galaxies. The center panel of Figure \ref{fig:NukerGals} shows the scalings of the specific tidal disruption rate, $\dot n$, which is plotted scaled to the specific disruption rate of solar type stars $\dot n_\odot$. Lines are colored here by the galaxies classification by  \citet{Faber:1997fn} as core or power-law, and we ignore collisions for simplicity. Where the tidal radius is less than the Schwarzschild radius, we still compute the $\dot n (\rt)$, but color the lines in gray. 

We find substantial variation between galaxies: a $100R_\odot$ star is between 2 and 10 times more likely to be disrupted than a solar-type star depending on the galaxy in which it resides. The dashed lines in the center panel of Figure \ref{fig:NukerGals} show power law scalings $\dot n \propto \rt^\alpha$ with $\alpha = 0,0.25,0.5,0.75,$ and $1$. Many galaxies fall in the range of $\alpha \sim 0.2 - 0.5$, broadly consistent with the  simplified model presented in Section \ref{sec:KepLC}, which found $\alpha = 1/4$. Another effect is a general trend toward shallower slopes $d {\rm ln} ({\dot n}) / d {\rm ln} (\rt)$ as $\rt$ increases, as seen in the right panel of Figure \ref{fig:NukerGals}. The change in slope is most significant in core galaxies with massive black holes. The peak in loss cone flux moves out in the cluster as $\rt$ increases, feeding stars from regions of lower stellar number density. In core galaxies, the peak in $\Flc$ may cross $\rb$ with significant effect on the scaling of $\dot n$ arising from the change in slope of the stellar number density at the break radius. In general, we find that the tidal disruption rate appears to vary more from galaxy to galaxy than between core and power-law classes. In what follows, we will use the relatively weak scaling $\dot n \propto \rt^{1/4}$ as a value that is representative of the majority of galaxies, irrespective of their structure. 

\subsection{The stellar diet of SMBHs}\label{sec:TDSE}

We have calculated how the specific  tidal disruption rate  $\dot n$ changes as a function of $\rt$. The tidal radius of a star is a function of the initial stellar mass and age (see Figure \ref{fig:EvolveStages}), so one must integrate over a stellar lifetime to find the relative disruption probability for stars during different evolutionary stages. Because $\dot n$ is a cluster-integrated rate, we are implicitly assuming that stars of different evolutionary stages are distributed isotropically in energy space. 
The expectation value for the number of disruption events a particular star will experience during an evolutionary stage lasting from $t_1$ to $t_2$ is given by the integral of $\dot n(\rt)$ across that time period, 
\beq
n_{12}=\int_{t_1}^{t_2} \dot n\left(\rt \right)  dt,
\eeq
where $\rt$ is a function of  $M_\ast(t)$ and $R_\ast(t)$.
Similarly, the total lifetime integral is
\beq
n_{\rm tot}=\int_{t_{\rm ZAMS}}^{t_{\rm max}} \dot n\left(\rt \right)  dt,
\eeq
where $t_{\rm ZAMS}$ is the age of the star at the zero-age main sequence, and $t_{\rm max}$ is the age at the  end of the star's lifetime. 
Therefore, the fractional likelihood of disruption during a given evolutionary stage as a function of initial stellar mass is given by
\beq\label{fstage}
f_{\rm stage}(M_{\rm ZAMS}) = \frac{n_{12}}{n_{\rm tot}}.
\eeq
We can consider the probability of being disrupted at different evolutionary  stages to be divided  in this manner as long as most  stars  of a given $M_{\rm ZAMS}$ do not experience a disruption during  their  lifetime ($n_{\rm tot} <1$). The fact that we observe giant stars in our own galactic center is a testament to the fact that many stars must survive without being disrupted during their entire lifetime \citep{Genzel:2010jk}. This requirement is most likely to be satisfied for stars that are only weakly bound to the black hole (of which there are many and for which the black hole is an extremely small target). In this case, which corresponds to the peak in the feeding rate  of giant stars, a typical star can easily survive without being disrupted over its entire  lifetime, and we have $N/\dot N \gg t_{\rm max}$, indicating that $n_{\rm tot} \ll1$. However, our assumption is less justified when considering stars well inside the sphere of influence. Within this smaller reservoir of stars there is a relatively high probability of disruption integrated over the star's  lifetime. 
In this case, there is a non-negligible chance that the star would have already suffered a tidal interaction with the black hole before reaching a given evolutionary stage -- either by evolving onto the loss cone \citep{Syer:1999gp} or by the typical diffusion process. 

\begin{figure*}[tbp]
\begin{center}
\includegraphics[width=0.99\textwidth]{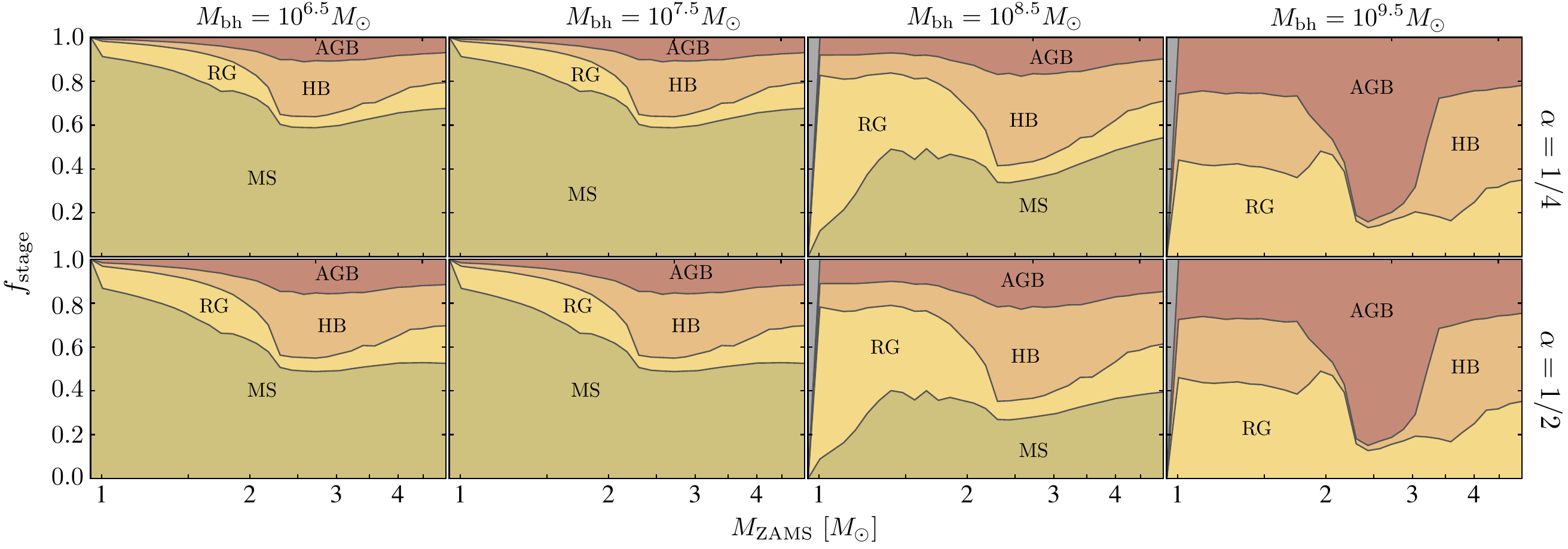}
\caption{The stellar ingestion flaring diet of SMBHs is shown for a range of SMBH masses, and two scalings of $\dot n$ with $\rt$, $\dot n \propto \rt^\alpha$.  The fractional likelihood of tidal disruption during different stellar evolutionary stages, $f_{\rm stage}$, is plotted as a function of initial stellar mass, $M_{\rm ZAMS}$. Stars more massive than about 1 solar mass evolve to the post-MS in less than a Hubble time and these evolved stars contribute a significant fraction of disruption events. As $\Mbh$ increases, an increasing fraction of MS stars are swallowed whole rather than being tidally disrupted. Flaring events on the most massive black holes consist exclusively of post-MS stars.   }
\label{fig:Menu}
\end{center}
\end{figure*}

The fractional likelihood of tidal disruption during different evolutionary stages, equation \ref{fstage}, is shown in Figure \ref{fig:Menu} for a range of initial stellar masses, $M_{\rm ZAMS}$. We show how this flaring diet depends on the SMBH mass and the scaling of $\dot n$ with $\rt$. In calculating $f_{\rm stage}$, we assume that the stellar population is old enough that stars are able to reach each of these representative evolutionary stages. The top panels in Figure \ref{fig:Menu}  show  the relative  fractions of disrupted stars at different evolutionary stages  given a $\dot n \propto \rt^{1/4}$ scaling, which we  found to be representative of most galactic centers. The bottom panels show the result given the optimistic scaling of $\dot n \propto \rt^{1/2}$. For black holes of relatively low mass (the sort that typically inhabit power-law galaxies), disrupted stars are typically MS stars with giant stars contributing 10-40\% of all disruption events. As black hole mass increases, an increasing fraction of MS stars are consumed whole rather than disrupted. The flaring diet of the most massive black holes consists nearly exclusively of post-MS stars. The disruption fractions corresponding to a population of stars can be constructed using the estimates of  $f_{\rm stage}$ given in Figure \ref{fig:Menu} by weighting them with the stellar population mass spectrum and age distribution.

%
%

\section{Discussion}\label{sec:disc}
We have shown that the tidal disruption of stars while  they are on the giant branch gives rise to longer-lasting accretion flares than  when disruption occurs during the MS  (Figure \ref{fig:DMDT_all}). Additionally, the stellar ingestion flaring diet of the most massive SMBHs consists exclusively of evolved stars (Figure \ref{fig:Menu}). The question remains whether these long duration accretion events occur frequently enough for  flaring activity to be a discernible characteristic  of the most massive SMBHs. In this section, we use Monte Carlo realizations of flaring events to illustrate the defining properties and expected rates of flaring events as a function of black hole mass, their contribution to the accretion luminosity of local AGN, and the prospects for detecting AGN flares powered by the disruption of evolved stars.

\subsection{A limiting SMBH mass scale for tidal disruption flares}

\begin{figure*}[tbp]
\begin{center}
\includegraphics[width=0.9\textwidth]{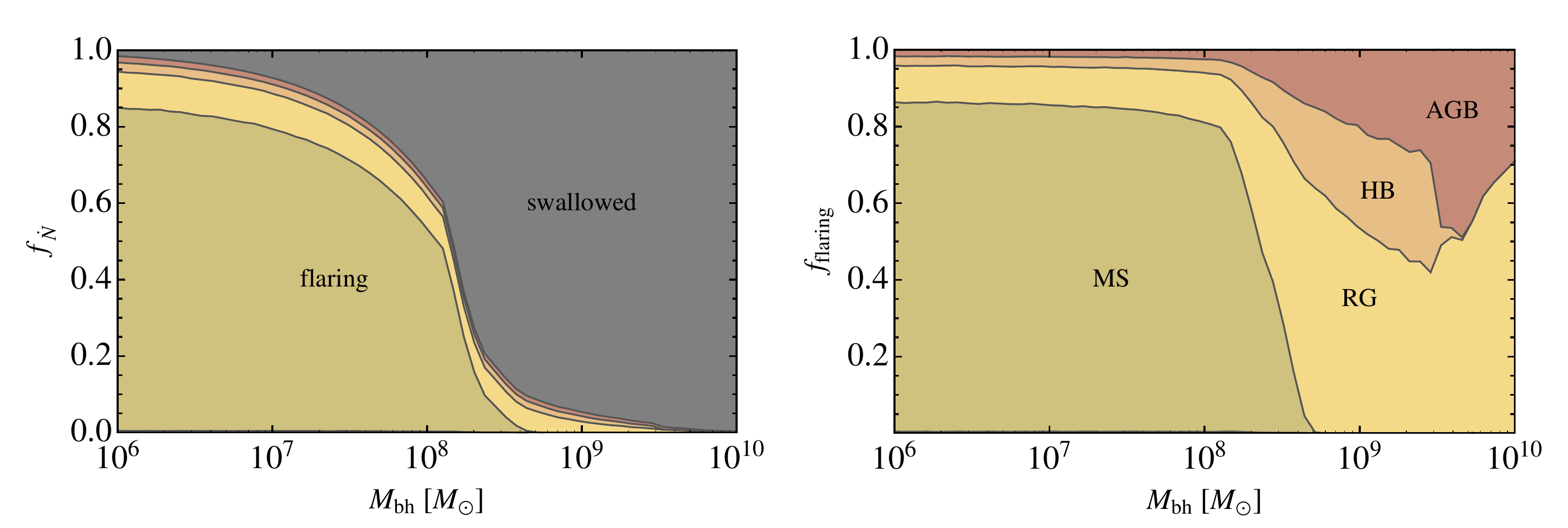}
\caption{In the left panel, we show the fractional composition of stars scattered into the loss cone, $f_{\dot N}$, for a range of black hole mass between $10^6 M_\odot$ and $10^{10} M_\odot$. We show a sharp cutoff in the flaring rate for black holes more massive than$10^8M_\odot$, beyond which MS stars are swallowed whole rather than tidally disrupted. Post-MS stars contribute some flaring events above this cutoff, but the weak scaling of $\dot n$ with tidal radius (Figure \ref{fig:FlcKepScalingRt}) dictates that these events are few compared to the flaring rate of lower mass black holes. The right panel shows the demographics of the flaring fraction only. These plots are based on a Monte Carlo realization of a single stellar mass $(1.4M_\odot)$ cluster with a flat distribution in stellar ages, and an $\dot n \propto \rt^{1/4}$ scaling law for the likelihood of individual events.}
\label{fig:FlareSwallow}
\end{center}
\end{figure*}

Giant stars are the only stars that can light up the most massive SMBHs by tidal disruption, since MS stars are swallowed whole rather than being tidally disrupted and subsequently producing a luminous accretion flare. Giant disruptions are, however, relatively infrequent due to the combination of the short lifetimes of evolved stars and the weak scaling of the tidal disruption rate with tidal radius $\dot n \propto \rt^{1/4}$. Therefore, despite their large radii, evolved stars are unable to compensate for the drop in flaring rate at black hole masses where MS stars are swallowed whole. As a result, we find a sharp cutoff in the tidal disruption flaring rate around $\Mbh \approx 10^8 M_\odot$.  This  is illustrated in Figure \ref{fig:FlareSwallow}, where, at low SMBH masses, the typical  event is a MS disruption with evolved star disruptions contributing about 15\%.  As the black hole mass grows, an increasing fraction of MS stars are swallowed whole rather than disrupted, eventually leading to the sharp cutoff in the flaring rate seen in Figure \ref{fig:FlareSwallow}.

\begin{figure*}[tbp]
\begin{center}
\includegraphics[width=0.99\textwidth]{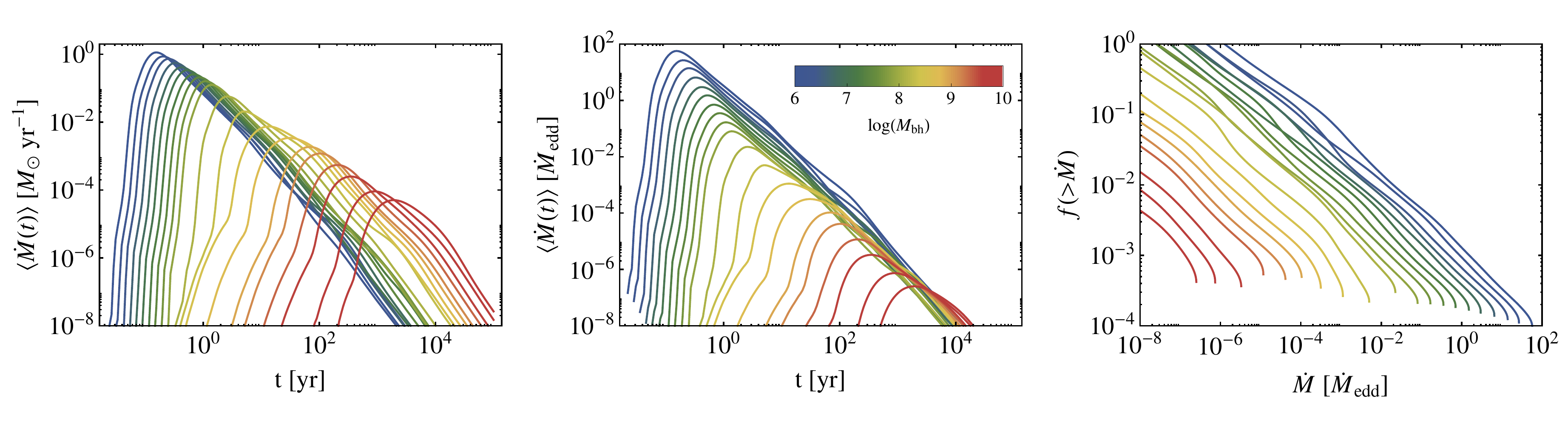}
\caption{The left panel and center panels show the average of Monte Carlo realized flaring events in physical units and as a ratio to the Eddington rate, respectively. For simplicity, we take all events to be $\beta = 1.5$ encounters of a $1.4 M_\odot$ ZAMS mass star. We draw events along the star's lifetime using an $\dot n \propto \rt^{1/4}$ scaling assuming a flat distribution of stellar ages in the nuclear cluster, then scale to the nearest of our simulations in Figure \ref{fig:DMDT_all}.  As black hole mass increases, the dominant flaring event goes from a MS disruption to a post-MS disruption, with the transition being particularly sharp around $\Mbh \sim 10^8 M \odot$. Post-MS disruptions feed gas to the black hole over longer durations than MS disruptions at correspondingly lower peak accretion rates. When normalized to the Eddington accretion rate of the black holes, $\dot M_{\rm edd} = 0.02 \epsilon_{0.1}^{-1} (\Mbh/10^6 M_\odot) \ M_\odot {\rm yr}^{-1}$, this amounts to a precipitous drop in the luminosity of tidal disruption powered SMBH flaring as black hole mass increases. The right panel shows the duty cycle of flaring luminosity, taking the overall loss cone flux from \citet{Wang:2004jy}, $\dot N \approx 6.5 \times 10^{-4} \left(\Mbh / 10^6 M_\odot\right)^{-1/4}$. Tidal disruption flares may be expected to contribute to the luminosity function of AGN harboring low mass black holes. However, they only contribute at very low luminosities relative to Eddington and with a low duty cycle in more massive SMBHs.  }
\label{fig:mcflaring}
\end{center}
\end{figure*}

The limiting black hole mass scale of $10^8 M_\odot$ coincides with a change in the SMBH flaring diet. In the right panel of Figure \ref{fig:FlareSwallow}, the typical disrupted star at high black hole masses is a giant branch star. At the very high mass end $\Mbh \gtrsim 10^{9.5} M_\odot$, and even HB stars are swallowed whole. The transition in the typical evolutionary state of disrupted stars as a function of black hole mass may also be inferred from Figure \ref{fig:mcflaring}, in which  the average tidal disruption mass feeding curve is plotted as a function of SMBH mass. For a population of SMBHs with $M_{\rm bh}\lesssim 10^{7}M_\odot$, the average $\dot M$ curve  is dominated by MS disruptions with some perturbation at late times due to disruptions of post-MS stars. The representative  $\dot M$ curves remain relatively unchanged until the black hole mass reaches the characteristic scale of $10^8 M_\odot$, where they transition rapidly to being  dominated by post-MS disruptions. 
While post-MS $\dot M$  curves peak near the Eddington rate for low mass black holes,  they fall  well below the Eddington rate for the most massive SMBHs, as illustrated in the center panel of Figure \ref{fig:mcflaring}. 
Supplying mass near the Eddington rate of black holes $\gtrsim 10^8 M_\odot$ via tidal disruption is not possible because of the transition to disruptions of stars exclusively in evolved stages.

Our finding of a limiting SMBH mass scale is at odds with the results of \citet{Syer:1999gp} who compute rates of giant consumption by SMBHs in Nuker galaxies and predict  a high rate of disruption of giant stars in the most massive galaxies. This is because \citet{Syer:1999gp} consider a process in which the evolution of closely approaching stars  results in mass transfer events with the black hole when they evolve off the MS.  Contrary to their assumption, however, this process results in many weak encounters with the black hole ($\beta \ll 1$) rather than a single disruptive encounter\footnote{The orbital period of trapped stars about the black hole, $P \sim 10^4$ years, is much less than even the post-MS stellar evolution timescale, $\tau_{\rm g} \sim 10^8$ years, resulting in very small per-orbit changes in stellar radius  and impact parameter, $\beta$.}. The sum of many encounters with the black hole may produce an interesting population of tidally-heated stars \citep{Alexander:2003jr,Alexander:2003dp}, but, is unlikely to produce luminous flares.

\subsection{Contribution of giant star disruptions to local low-luminosity AGN}

Having established a limiting SMBH mass scale for tidal disruption flaring events, we examine how this mass limit manifests itself in the cumulative, low-luminosity output of local AGN. 
Tidal disruption flaring events are observed to rise sharply  and then decay over long timescales approximately following the widely discussed $t^{-5/3}$ power law. The contribution of the tidal disruption of solar-type stars to the local AGN luminosity function was investigated by \citet{Milosavljevic:2006jj} who found that the long decay tails of tidal disruptions can contribute to the integrated luminosity of AGN in gas-starved galaxies. The exact value of the  tidal disruption rate as well as its evolution with redshift \citep[which is rather uncertain; although see][]{Merritt:2009ir}  determines the time at which gas-mode feeding of active galaxies drops to sufficiently low levels for the stellar disruption component to take over \citep[e.g.][]{Heckman:2004js}. For example, \citet{Milosavljevic:2006jj} find that with a tidal disruption event of $\dot N \approx 6.5 \times 10^{-4} \left(\Mbh / 10^6 M_\odot\right)^{-1/4}$, the associated flaring luminosity is likely to be responsible for at least 10\% of local ($z\leq 0.2$) AGN activity.

Observational AGN surveys have emphasized a picture of cosmic downsizing of black hole growth. The most massive SMBHs grew during the quasar era \citep{Yu:2002fx}, while the black holes believed to be actively growing now are much less massive, around $10^7 M_\odot$ \citep{Heckman:2004js}.  Our finding of a limiting mass scale for tidal disruption powered flaring is consistent with the observation that most local, massive SMBHs are quiescent. Tantalizingly, the distribution of local AGN-hosting galaxies is biased to about an order of magnitude lower SMBH mass than the general galaxy population \citep{Greene:2007dz}, with a decrease in activity around $\Mbh \approx 10^{8.5} M_\odot$ that lies coincident with the limiting SMBH mass scale for tidal disruption flares. 

The right panel of Figure \ref{fig:mcflaring} examines the duty cycle of tidal disruption powered flaring in local AGN. 
Despite their lower frequency of disruption, giant star flares make a sizable contribution  to the average duty cycle at low luminosities because their power law decay tails have higher normalization than those of MS stars (see Figure \ref{fig:DMDT_all}).   The combined effects of the decrease in stellar consumption rate \citep{Wang:2004jy}, the precipitous drop in the flaring fraction (Figure \ref{fig:FlareSwallow}) and the change in the demographics of disrupted stars (Figures \ref{fig:FlareSwallow} and \ref{fig:mcflaring}) all conspire to ensure that tidal disruption powered flares contribute well below Eddington and with a very low duty cycle (a small number in the active state) on the most massive SMBHs.

\begin{figure}[tbp]
\begin{center}
\includegraphics[width=0.35\textwidth]{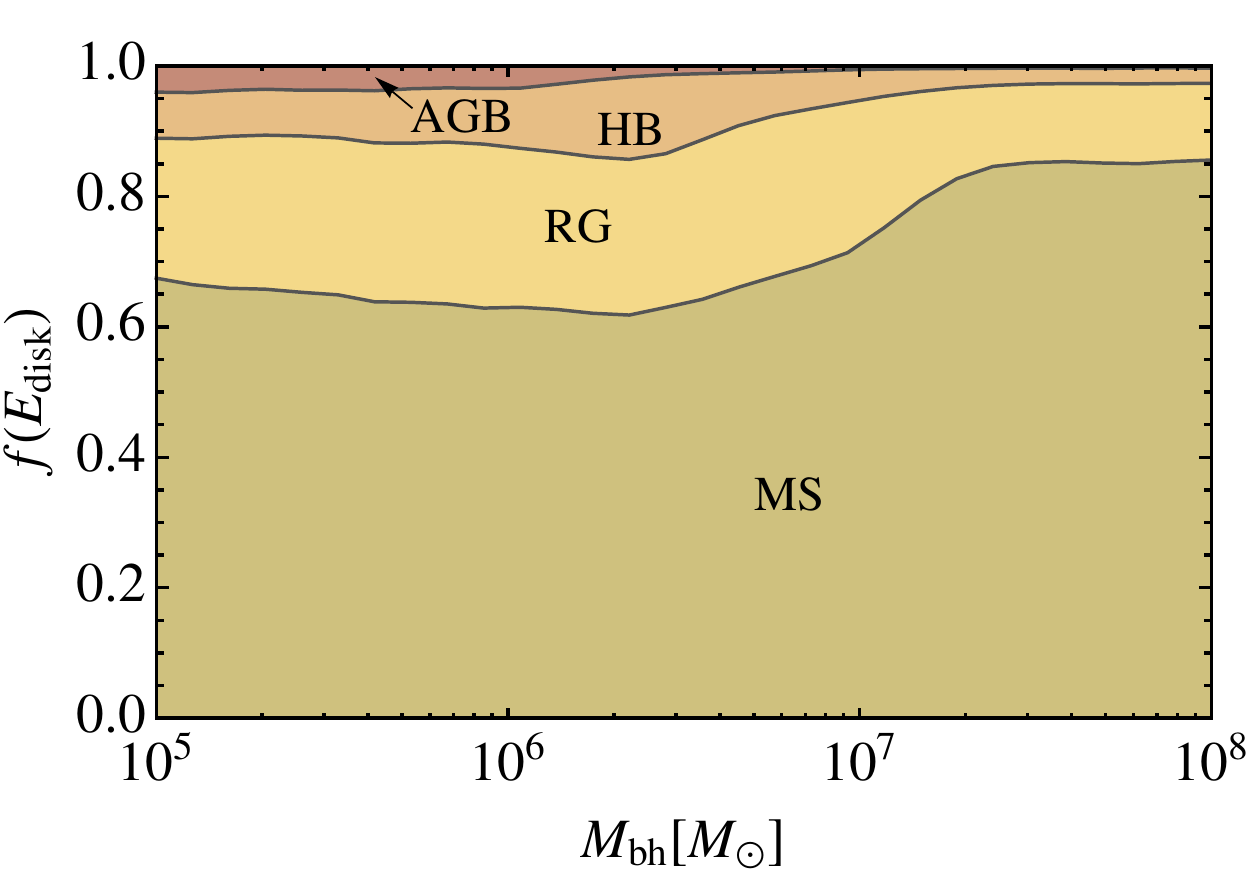}
\caption{The fractional distribution of total tidal disruption fed accretion disk light emitted, $E_{\rm disk}$, calculated by a Monte Carlo realization of the mass fed to SMBHs below their Eddington limit, again for a $1.4M_\odot$ star and $\beta = 1.5$ encounters. At lower SMBH masses, MS disruptions peak at highly super Eddington black hole feeding rates. As a result, MS disruptions inefficiently convert stellar mass loss into accretion luminosity. However, giant stars feed more of their mass to SMBHs below the Eddington limit, resulting in a more efficient conversion to disk luminosity. Thus, while post-MS disruptions occupy about 15\% of disruption events by rate, in terms of total disk emission, they contribute nearly 40\% to the ensemble. As the SMBH mass increases to $10^8 M_\odot$, both MS and post-MS disruptions are sub-Eddington, and the fractional contributions to $E_{\rm disk}$ are similar to the event rate convolved with the mass loss per event (see Figures \ref{fig:FlareSwallow} and \ref{fig:DM_all}, respectively).} 
\label{fig:Edisk}
\end{center}
\end{figure}

The relative contributions of MS and evolved stars to the cumulative, tidal disruption-fueled accretion disk luminosity of local AGN depends on the total amount of stellar mass  fed to SMBHs at rates below their corresponding Eddington limits. 
Super-Eddington fallback results in an inefficient conversion of fallback mass into light, since the accretion rate onto the black hole (and, correspondingly, the accretion disk luminosity) is capped at or near the Eddington limit\footnote{The super-Eddington fallback phase may, however, give rise to outflows (rather than accretion of mass with low radiative efficiency)  and a conversion of mass to light at larger radius as the ejected debris expands \citep{Strubbe:2009ek,Strubbe:2011iw,Kasen:2010ci} }.
Because their disruption results in highly super-Eddington feeding of lower mass SMBHs around $10^6 M_\odot$, MS disruptions are rather inefficient at converting mass stripped from the disrupted stars into accretion luminosity.
Giant stars disruptions, on the other hand, peak at lower accretion rates and feed more of their mass to the black hole below the Eddington limit. 
Figure \ref{fig:Edisk} shows the fractions of accretion disk luminosity contributed by stars at different evolutionary states. Giant star disruptions make up for in efficiency what they lose in disruption frequency when compared to MS stars and, thus, contribute  significantly  to the overall power of tidal disruption-fueled AGN. 


\subsection{Detection of giant star tidal disruption flares}

\begin{figure}[tbp]
\begin{center}
\includegraphics[width=0.4\textwidth]{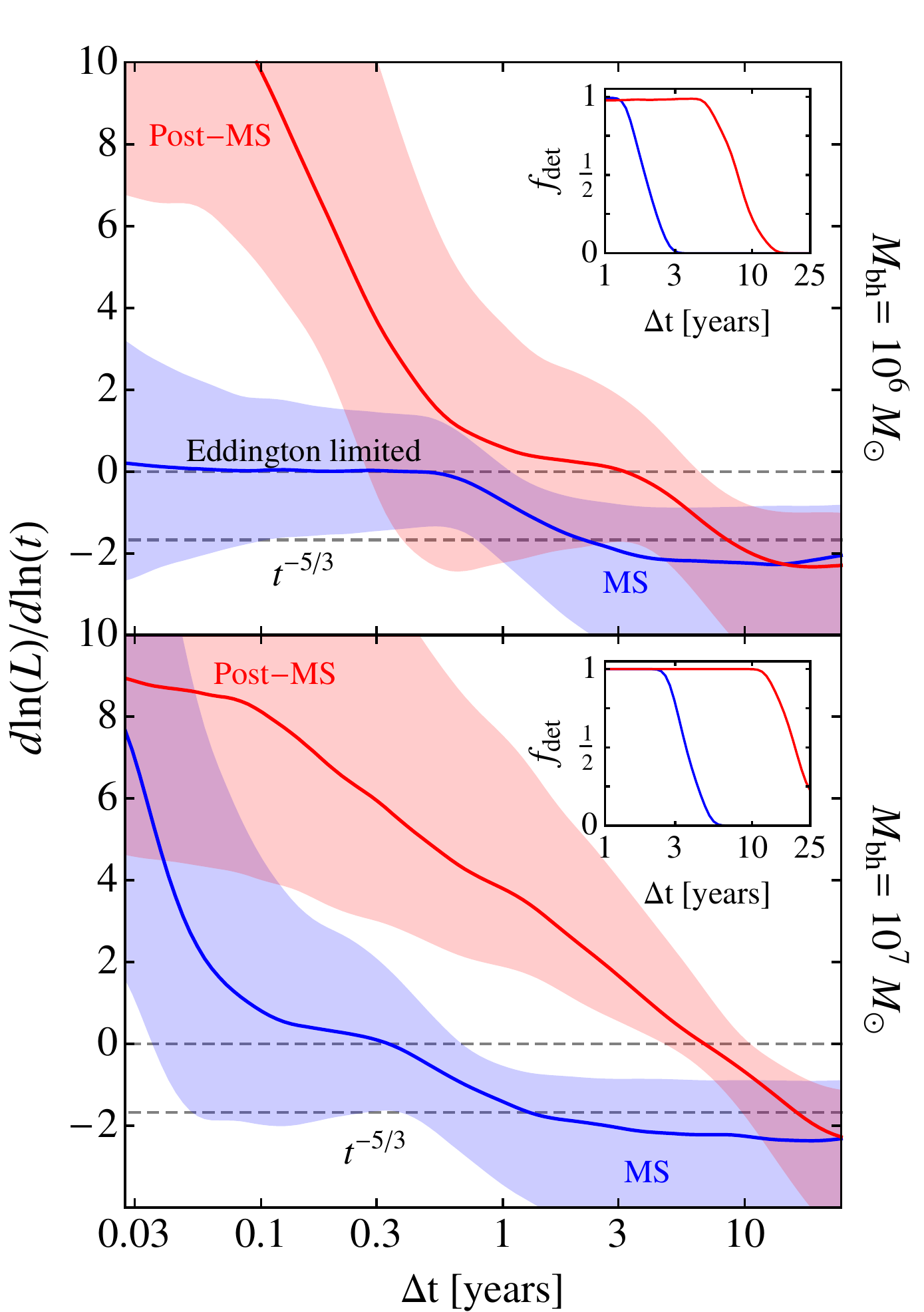}
\caption{ The power-law slope inferred from two successive detections of hypothetical MS and post-MS tidal disruption flare transients given different observing cadences. Observing cadences that initially detect MS disruption events in the plateau or decay phases will catch the rise to peak in post-MS disruption flares. This plot assumes that flares above an intrinsic luminosity of $10^{43} \ \text{erg s}^{-1}$ may be detected. This is roughly equivalent to the bolometric luminosity at which the recent tidal disruption flare PS1-10jh at $z=0.1696$ was initially detected \citep{Gezari:2012fk}. At the Eddington limit, we assume the luminosity of these flares plateaus but is convolved with a normally distributed variability factor with $\sigma = 0.01 (\Delta t/1 {\rm yr})^{1/2}$. The flares are all assumed to occur at an impact parameter of $\beta =1.5$. The post-MS curve is the average of the RGI and HB models (see Figure \ref{fig:DMDT_all}). In both cases, the lines plotted are the median while the shading shows the $\pm 1\sigma$ region for a Monte Carlo realization of detection times relative to the observing cadence.  The inset plots show the detectable fraction of events, $f_{\rm det}$, defined as those typically visible with at least two observations above the $10^{43} \ \text{erg s}^{-1}$ threshold given the cadence.  } 
\label{fig:cadence}
\end{center}
\end{figure}

The primary characteristic that distinguishes the tidal disruption flares of evolved  stars from  those of MS stars is their comparatively long timescales. 
Tidal disruption events are typically characterized in transient surveys by their late-time $\sim t^{-5/3}$ decay \citep[see][for some recent examples]{Gezari:2009dna,vanVelzen:2011gz,Saxton:2012wv}. However, as seen in Figure \ref{fig:DMDT_all}, We are unlikely to observe the characteristic power-law decay in the evolved cases due to their extremely long characteristic decay times. Our simulations show that the rise up to peak luminosity in giant star flares occurs over similar timescales as the decay in MS disruption flares. 
Figure \ref{fig:cadence} illustrates that typical observing cadences will detect giant star disruptions during  their rise to peak brightness. Shown in the post-MS category in  Figure \ref{fig:cadence} are the average mass feeding rates of the RG I and HB disruption simulations. Tidal disruptions of stars at the tips of the giant branches (the RG II and AGB models) result in flares with such long timescales that they would be difficult to identify in transient surveys. The power law slope during the rise phase  is quite steep, indicating that the brightness of a galactic nucleus would increase in a correlated fashion by many orders of magnitude over a period of a few years. 

AGN are infamous for exhibiting variability at a range of timescales, and one might worry that this would frustrate the detection of giant disruption flares. However, the power spectrum of these variations is well below the amplitude  expected from  the disruption of a post-MS star. 
\Citet{vanVelzen:2011gz} perform a detailed analysis of potential sources of tidal disruption event confusion, including an empirical analysis of the amplitude of AGN flaring events in their Sloan Stripe 82 sample of 1304 AGN. They find that the largest amplitude flaring events exhibit changes in flux of $\Delta F/ F \lesssim 5$, far less than the amplitude of the correlated rise expected from the disruption of an evolved star. More detailed studies of AGN variability also find small correlated increases \citep{Kelly:2009fx,Kelly:2011ku,MacLeod:2010ex}, giving credence to the idea that the tidal disruption of an evolved star would imprint  a unique signal in the AGN flaring history.

The prospects for detecting giant star flares in the near future are, in fact, quite promising. 
While these events are less common than MS disruptions, they should be commonly detected in future transient surveys, like {\it LSST}, which are expected to yield hundreds of MS disruption events per year \citep{Strubbe:2009ek}.
The insets in Figure \ref{fig:cadence} show that finding giant star disruption events does not require short cadence surveys. 
It is even possible that the signatures of giant star disruptions are lurking in current data sets and may be detectable through binning the data at lower cadences. Tidal disruption events with well sampled light curves that facilitate detailed comparison with theoretical $\dot M$ profiles offer the best hope of characterizing the physics of tidal disruption. 

\subsection{Caveats and  prospects}

In this work we have assumed that when the tidal radius is less than the Schwarzschild radius stars are swallowed whole without producing a flare. The impact parameter at which this transition occurs, however, 
is likely sensitive to the orientation of the encounter relative to the spin of the black hole \citep{Kesden:2012cn,Haas:2012ci}.  The exact fraction of events which are promptly swallowed or produce flares thus depends on a general relativistic description of the black hole's potential.

A great deal of uncertainty lies in the properties of the nuclear star clusters from which stars are fed into disruptive orbits. 
We know very little about stellar kinematics  in galactic centers other than our own \citep[e.g.][]{Genzel:2010jk}. Our work has made the usual assumptions of a spherical nuclear star cluster which feeds stars to the black hole by a  two-body relaxation-driven random walk in angular momentum space. Theoretical work has emphasized a wide variety of effects which may enhance the rate at which stars are fed towards the black hole. A triaxial potential can feed stars to the loss cone collisionlessly through chaotic, boxy orbits \citep{Magorrian:1999fd,Merritt:2004jc,Merritt:2010he}. Rings or disks of stars, if present, would feed stars to the black hole at an enhanced rate through secular instabilities \citep{Madigan2011}. Another secular effect is resonant relaxation, which almost certainly enhances the loss cone flux to some extent for the most bound stars \citep{Magorrian:1999fd,Hopman:2006bq,Gurkan:2007bv}. Finally, a second massive body, like an insprialing black hole or a giant molecular cloud could easily induce large angle scatterings of stars   \citep{Ivanov:2005io,Perets:2007fo,Chen:2008gr,Chen:2009bt,Chen:2011cs,Wegg:2010ue}. All of these processes result in more completely filled loss cone  than the conservative scenario we've outlined here.

Further, even the simplest characteristics of galactic center stellar populations, like the age and mass distributions of stars, are wildly uncertain. Stars are recycled from the galaxy in general as the interaction between tightly bound stars and the more weakly bound galactic stellar population lead to a time evolution of the nuclear cluster compactness and properties \citep{Merritt:2009ir}. In fact, our own galactic center is marked by a striking and enigmatic population of massive, young stars in the central $\sim 0.1$ pc \citep[e.g.][]{Schodel2007} that lies in contrast to the underlying, old stellar population. There are various alternative lines of evidence  further suggesting that the initial mass function (IMF) of stars   surrounding our own galactic center may be very different from that representative of the bulk of the star formation \citep{Maness:2007gm,Bartko:2010hh}. 

The dynamics of a cluster comprised of a spectrum of stellar masses is also expected to operate somewhat differently than that of a single mass cluster. In power law galactic centers, which are presumed to be dynamically relaxed \citep{Bahcall:1976kk,Bahcall:1977ea}, one expects that massive stars, binaries, and stellar remnants to segregate to tightly bound orbits over the cluster's relaxation time. An illustrative example of these effects comes comes from simulations of globular clusters harboring intermediate mass black holes (IMBHs), which have been performed with direct N-body integrations \citep[e.g.][]{Baumgardt:2004jf,Baumgardt:2004dx,Trenti:2007ie} and Monte Carlo techniques \citep[e.g.][]{Gurkan:2004et}. For example, \citet{Baumgardt:2004dx} consider a multi-mass stellar cluster and find that the massive stars segregate to form a tightly bound cusp around the IMBH, while the less massive stars relax to a shallower distribution.  The scaling of the specific tidal disruption rate, $\dot n$,  is modified  by the varying radial profile from which stars are fed. In this case, more massive stars are fed into the loss cone preferentially as compared to low mass stars, and the rates of giant star disruption in their simulations are 15-20\% of the total stellar disruption rate, a factor of $\sim 2$ greater than one derives from a simple weighting over the stellar mass spectrum. 

The detection of tidal disruptions holds great promise for revealing the properties of quiescent SMBHs and the nuclear star clusters that surround them, which, as we have discussed, are rather unconstrained. The universe is now being explored in a panchromatic way over a range of temporal scales, leading toward a more complete and less biased understanding of its constituents. Flares from the disruption of stars in all evolutionary states will certainly be observed in the not-too-distant future. The recent observations of PS1-10jh  presented by \citet{Gezari:2012fk} offer a glimpse into disruptions of stars not on the MS.   Taken in a statistical sense, the observed rates of tidal disruption and, in particular, the relative rates of disruptions of different stellar evolutionary stages, will hold tremendous distinguishing power in both the dynamical mechanisms typically operating in galactic centers and the properties of the populations of stars themselves. Will the most common tidal disruption flares originate from young, massive stars in disks? Or from old stars fed from the near the edge of the black hole's sphere of influence? It is within the capacity of planned surveys to obtain lightcurves of giant star tidal disruption flares that are well resolved temporally. These lightcurves will challenge our understanding of the physics of tidal disruption, and through comparison with simulated events, offer a window into the extreme nature of close encounters between stars and SMBHs.

\acknowledgments{
It is a pleasure to thank Suvi Gezari, Paul Groot, Dan Kasen, Stefanie Komossa, Ryan O'Leary, William Lee, Doug Lin, Shangfei Liu, Ann-Marie Madigan, Martin Rees, Stephan Rosswog, and Ken Shen for insightful comments and helpful discussions. The software used in this work was in part developed by the DOE-supported ASCI/Alliance Center for Astrophysical Thermonuclear Flashes at the University of Chicago. We acknowledge support from the David and Lucile Packard Foundation, NSF grant: AST-0847563, the NSF graduate research fellowship (M.M.), and the NESSF graduate fellowship (J.F.G.). }

\bibliographystyle{apj}


\end{document}